\documentclass[twocolumn]{pasj00}
\SetRunningHead{Astronomical Society of Japan}
{Lapse of X-ray eclipse in GRS~1747$-$312}

\usepackage{ulem}
\usepackage{color}

\begin{document}

\title{Peculiar Lapse of Periodic Eclipsing Event at Low Mass X-ray Binary GRS~1747$-$312 during Suzaku Observation in 2009}

\author{Shigetaka \textsc{Saji},\altaffilmark{1}
Hideyuki \textsc{Mori},\altaffilmark{2,}\altaffilmark{3}
Hironori \textsc{Matsumoto},\altaffilmark{4}
Tadayasu \textsc{Dotani},\altaffilmark{5, 6, 7}
Masachika \textsc{Iwai},\altaffilmark{6, 5}\\
Yoshitomo \textsc{Maeda},\altaffilmark{5, 7}
Ikuyuki \textsc{Mitsuishi}, \altaffilmark{1}
Masanobu \textsc{Ozaki},\altaffilmark{5}
Yuzuru \textsc{Tawara}, \altaffilmark{1}}

\altaffiltext{1}{Division of Particle and Astrophysical Science,
Graduate School of Science, Nagoya University, \\
Furo-cho, Chikusa-ku, Nagoya, 464-8602}
\altaffiltext{2}{CRESST and X-ray Astrophysics Laboratory, Code 602, NASA/Goddard Space Flight Center,
Greenbelt, MD 20771, USA}
\altaffiltext{3}{Department of Physics, University of Maryland, Baltimore County, 1000 Hilltop Circle, Baltimore, MD 21250, USA}
\altaffiltext{4}{Kobayashi-Maskawa Institute for the Origin of
Particles and the Universe, Nagoya University, \\
Furo-cho, Chikusa-ku, Nagoya, 464-8602}
\altaffiltext{5}{
Institute of Space and Astronautical Science (ISAS), Japan Aerospace
Exploration Agency (JAXA), \\
3-1-1, Yoshinodai, Chuo-ku, Sagamihara, 252-5210}
\altaffiltext{6}{Tokyo Institute of Technology, 12–1, Ookayama 2–chome, Meguro-ku, Tokyo, 152–8550, Japan}
\altaffiltext{7}{Department of Space and Astronautical Science, SOKENDAI (The Graduate University for Advanced Studies), 3-1-1 Yoshinodai, Chuo-ku, Sagamihara, Kanagawa 252-5210, Japan}


\email{s\_saji@u.phys.nagoya-u.ac.jp}

\maketitle

\begin{abstract}
GRS~1747$-$312 is a neutron star Low-Mass X-ray Binary 
in the globular cluster Terzan~6, located at a distance of
$9.5$~kpc from the Earth. 
During its outbursts, periodic eclipses were known to occur. 
Observations for the outbursts were performed with Chandra in
2004 and Swift in 2013. XMM-Newton observed its quiescent
state in 2004. In addition, when Suzaku observed it in 2009 as
a part of Galactic center mapping observations, 
GRS 1747$-$312 was found to be in 
a low luminosity state with $L_{\rm x} \sim 1.2 \times 10^{35}$ erg s$^{-1}$.
 All of the observations except for XMM-Newton included the time of the eclipses predicted. 
We analyzed archival data of these observations. 
During the Chandra and Swift observations, we found clear flux
decreases at the expected time of the eclipses. During the
Suzaku observation, however, there were no clear signs for the predicted eclipses. 
The lapse of the predicted eclipses during the Suzaku observation
can be explained by a contaminant source quite close to GRS~1747$-$312.
When GRS~1747$-$312 is in the quiescent state, we observe X-rays from
the contaminant source rather than from GRS~1747$-$312.
However, we have no clear evidence for the contaminant source in our data.
The lapse might also be explained by thick material ($N_{\rm H} > 10^{24}$~cm$^{-2}$ )
between the neutron star and the companion star, 
though the origin of the thick material is not clear.
\end{abstract}

\section{Introduction}

Low Mass X-ray Binaries (LMXBs) are stellar systems
consisting of a compact object and a low-mass ($\lesssim
1M_\odot$) companion star.  The companion star supplies gas
onto the compact star by Roche-lobe overflow. The accreting
matter releases the gravitational energy and then emits
X-rays. 
Many LMXBs exhibit large variability in X-ray luminosity 
between $10^{32} - 10^{33}$~erg~s$^{-1}$ (quiescent state) 
and $10^{37} - 10^{38}$~erg~s$^{-1}$ (outburst state).
So far, more than $100$ LMXBs have been identified in the
Galaxy~\citep{2007A&A...469..807L}. Among them, only six are known to have clear X-ray
eclipses: EXO~0748$-$676 \citep{2009ApJS..183..156W},
XTE~J1710$-$281 \citep{2011MNRAS.413....2J}, X~1658$-$298
\citep{2001A&A...376..532O}, AX~J1745.6$-$2910
(\cite{1996PASJ...48..417M}; \cite{1996PASJ...48L.117K}),
4U~2129$+$47 \citep{2009ApJ...706.1069L} and GRS~1747$-$312
(this work). An eclipsing feature informs us of an accurate
orbital period and a strong limitation on an inclination
angle of the system.

GRS~1747$-$312 is a transient X-ray source firstly detected
in 1990 with ROSAT \citep{1991A&A...246L..21P} and Granat
\citep{1994ApJ...425..110P}.  The source is located at
$(\alpha, \delta)_{\rm J2000.0} $ =
($\timeform{267.69526D}$, $\timeform{-31.27468D}$) or $(l,b)
= (\timeform{358.57274D}, \timeform{-2.16286D})$. 
The position was determined with the Chandra
HRC-I with an uncertainty of $\timeform{0."4}$
\citep{2003A&A...406..233I}.  The position is coincident
with Terzan~6, which is a core-collapsed metal-rich globular
cluster \citep{1997A&AS..122..483B}. The distance to the
cluster is estimated to be $9.5^{+3.3}_{-2.5}$~kpc from the
luminosity level of the horizontal
branch~\citep{2003A&A...399..663K}.
The compact object in GRS~1747$-$312 is considered to be a neutron star,
since four type-I X-ray bursts from GRS~1747$-$312 were detected \citep{2003A&A...406..233I}. Assuming the distance for Terzan~6, the peak flux of the type-I bursts was consistent with the flux of a neutron star at the Eddington luminosity. Thus GRS~1747$-$312 is thought to belong to 
the cluster. In the HRC-I observation, no other X-ray sources
were detected in the cluster. The detection limit of the
observation was $1.3 \times 10^{-13}$~erg~cm$^{-2}$~s$^{-1}$~\citep{2003A&A...406..233I}.

It is known that GRS~1747$-$312 exhibits recurrent X-ray
outbursts with an interval of about
$130$--$142$~days~\citep{2003A&A...406..233I}. Typical time scale of the decay of the outbursts was $\sim18$~days.
During the outburst, BeppoSAX and Rossi X-ray Timing Explorer (RXTE) observations found
periodic eclipses with a duration of $2596$~s.  The authors
claimed that the orbital period is $P = 0.514980303$~days.
The duration of the eclipse and the orbital period indicate
that the mass of the companion star is at least
$0.1$~M$_\odot$ assuming a mass of $1.4$~M$_\odot$ for the
neutron star. Based on the typical age of a globular
cluster, the mass of the companion star is thought to be
lower than $0.8$~M$_\odot$~\citep{2000A&A...355..145I}.

An energy spectrum of GRS~1747$-$312 during the outburst
obtained with BeppoSAX was investigated in
\citet{2000A&A...355..145I}.  No line-like structures were
detected in the energy range between $1$~keV and $100$~keV.
The spectrum was represented well by a two-component model
consisting of a blackbody and a Comptonized emission; the
blackbody temperature was $kT = 1.78$~keV, while the Comptonized
component has an electron temperature of $kT_{\rm e} =
5.4$~keV, an optical depth of $\tau = 7.4$, and a
temperature of seed emission of $kT_{\rm seed} = 0.57$~keV.
The spectral analysis demonstrated that the absorption
column density was approximately $10^{22}$~cm$^{-2}$, which
is consistent with the value obtained from the interstellar
reddening of Terzan~6 
\citep{2000A&A...355..145I, 1995A&A...293..889P}.

Previous studies of GRS~1747$-$312, however, were
concentrated on the outburst state.  Thus the behavior
outside the outbursts is still unclear.  In this paper, we
analyzed archival data of Chandra, XMM-Newton, Suzaku and
Swift to investigate physical properties in various states.
Detailed information on the observations is described in
section~\ref{section:observation_and_reduction}. Results on
imaging, timing and spectral analysis are reported in
section~\ref{section:analysis}. Discussions and
conclusions are given in
sections~\ref{section:discussion} and \ref{section:summary}, respectively.
In this paper, the uncertainties are at the 90\%
confidence level, while the errors on the data points 
of spectra, radial profiles, and light curves are at the $1 \sigma$ 
level, unless otherwise stated.

\section{Observation and data reduction
\label{section:observation_and_reduction}
}

\subsection{Chandra}

Chandra \citep{2000SPIE.4012....2W} has an X-ray telescope
named High Resolution Mirror Assembly with an angular
resolution of $\timeform{0."5}$.  The telescope focuses
X-rays onto one of two kinds of instruments, Advanced CCD
Imaging Spectrometer (ACIS; \cite{2003SPIE.4851...28G}) or
High Resolution Camera (HRC; \cite{2000SPIE.4012...68M}).
Chandra is equipped with two grating spectrometers, High
Energy Transmission Grating (HETG;
\cite{2005PASP..117.1144C}) and Low Energy Transmission
Grating (LETG; \cite{2000ApJ...530L.111B}); the HETG covers
the energy range of $0.4$--$10$~keV and the LETG covers
$0.07$--$0.2$~keV.

GRS~1747$-$312 was observed twice with Chandra on 2000
March~9 (ObsID $=$ 720) and on 2004 March~29 (ObsID $=$
4551).  The observation in 2000 was published by
\citet{2003A&A...406..233I} to determine the source position
with the HRC. Thus, we focus on the data obtained from the
observation in 2004 hereafter. The observation was conducted
as a Target of Opportunity (ToO) to obtain the data during
the outburst, and it covered the time of the eclipse (see
section~\ref{subsection:light_curve}).  The log of the
observation is summarized in
table~\ref{table:observation_log}.

The observation was operated with the ACIS-S in the
Continuous Clocking (CC) mode.  The ACIS-S contains six
X-ray CCD chips arranged in a $1\times6$ array, and each
chip has $1024\times1024$~pixels.  The field of view is
$\timeform{8.'3}\times\timeform{50.'6}$.  In the CC mode,
two-dimensional images cannot be obtained, but a high time
resolution of $2.85$~ms can be achieved.  An energy
resolution is $95$~eV at $1.49$~keV and $150$~eV at
$5.9$~keV.  During the observation, the HETG was inserted
between the telescope and the ACIS-S.  We only used the
zeroth order image for our analysis.

We analyzed the primary event data processed with
\texttt{Standard Data Processing pipeline Version DS 8.4}
with a software package \texttt{CIAO version 4.6} and a
relevant calibration database (\texttt{CALDB version
  4.6.1.1}).


\subsection{XMM-Newton}

GRS~1747$-$312 (a.k.a. 1RXS J175046.5$-$311632) was observed
with XMM-Newton in 2004 September 28.  XMM-Newton
\citep{2001A&A...365L...1J} enables us to conduct imaging
spectroscopy in the $0.3$--$10$~keV band, utilizing the
X-ray telescopes \citep{2000SPIE.4012..731A} and the focal
plane detectors called European Photon Imaging Camera
(EPIC).  EPIC consists of two types of X-ray CCDs: Metal
Oxide Semiconductor (MOS)  CCDs \citep{2001A&A...365L..27T}
and pn CCDs \citep{2001A&A...365L..18S}.  There are two MOS
detectors (MOS1 and MOS2). 
The EPIC cameras were operated with the medium filter in the
full-frame mode in this observation. 
The total exposure time were
$8.2$~ks for pn and $14.3$~ks for each MOS detector.  

We reprocessed the Observation Data Files provided by the
XMM-Newton Science Archive with \texttt{epchain} and
\texttt{emchain}, using appropriate calibration data.  We
then used the Science Analysis Software (\texttt{SAS, version 13.0})
to make images, light curves and spectra. In order to excise flaring
particle backgrounds mainly caused by particles such as soft
protons, we filtered out the reprocessed event files,
following the instruction in the XMM data reduction
guide\footnote{http://xmm-tools.cosmos.esa.int/external/xmm\_user\_support/\\documentation/sas\_usg/USG.pdf}.
We made X-ray light curves of single events
(\texttt{PATTERN} = 0) in the $10$--$12$~keV band for pn and
above $10$~keV for MOS1/2.  The events were accumulated from
the entire field of view of each detector.  The good time
intervals were selected so that the count rates are smaller
than $0.4$~cts~s$^{-1}$ for pn or $0.35$~cts~s$^{-1}$ for
MOS1/2.  We found that the observation was little affected
by the background flares: the screened live times for pn and
MOS1/2 were $8.2$~ks and $14.2$~ks, respectively. In the
following analysis, we used the screened event data. The
observation log is summarized in
table~\ref{table:observation_log}.


\subsection{Suzaku}

GRS~1747$-$312 was also observed with Suzaku
\citep{2007PASJ...59S...1M} in 2009 September 16. The
observation was performed as a part of the mapping
observation of the Galactic center region.  The information
on the Suzaku observation is also given in
table~\ref{table:observation_log}.  

One of the main instruments of Suzaku is the X-ray Imaging
Spectrometer (XIS) consists of four sets of CCD cameras
(XIS0-3) \citep{2007PASJ...59S..23K}.  Each CCD chip is
placed on the focal plane of the X-Ray Telescope (XRT;
\cite{2007PASJ...59S...9S}). XIS0, 2 and 3 are
front-illuminated (FI) CCDs, while XIS1 is a
back-illuminated (BI) CCD. Since XIS2 suffered serious
damage, it has not been usable since 2006 November 9.  In
this observation, the XIS was operated in the normal
clocking mode with the Spaced-row Charge Injection (SCI;
\cite{2009PASJ...61S...9U}).  Suzaku also has the Hard X-ray
Detector (HXD ; \cite{2007PASJ...59S..35T}).  The HXD
consists of Si-PIN photo diodes and GSO scintillators.
Although the HXD is a non-imaging detector, the PIN and GSO
provide us X-ray spectra covering the energy ranges of
$10$--$70$~keV and $40$--$600$~keV, respectively.  The HXD
was operated in the normal mode in this observation.

We analyzed the cleaned event data processed with the
\texttt{processing version of 2.4.12.27}.  In this process, the data
taken at the South Atlantic Anomaly (SAA) and low elevation
angles of $< 5$~deg from the night-earth rim and $< 20$~deg
from the day-earth rim were excluded. The net exposure time
was $45.3$~ks for the XIS and $45.6$~ks for the HXD.  To
analyze the data, we used the software package
\texttt{HEAsoft version 6.12}, 
and the relevant calibration database (\texttt{CALDB}).  We
generated the redistribution matrix files (RMFs) and
ancillary response files (ARFs) by \texttt{xisrmfgen} and
\texttt{xissimarfgen} \citep{2007PASJ...59S.113I}.

\subsection{Swift}

On 2013 March 11, INTEGRAL/JEM-X~\citep{2003A&A...411L.231L} detected a burst from
GRS~1747$-$312 \citep{2013ATel.4883....1C}.  Soon after the
detection, Monitor of All-sky X-ray Image (MAXI; \cite{2009PASJ...61..999M}) on board the
International Space Station (ISS) gave a transient alert on
March~11\footnote{http://maxi.riken.jp/pipermail/x-ray-star/2013-March/000210.html}.
Subsequently, two ToO observations were performed with Swift
on March~18 and March~24. The exposure for each observation
was $\sim 1$~ks.  
The 2nd observation was planned to observe the
predicted ingress of the eclipse.

We used Swift X-Ray Telescope (XRT) data of these
observations.  Swift XRT \citep{2004ApJ...611.1005G} consists
of an X-ray telescope and a single X-ray CCD
($600\times602$~pixel), which covers the $0.2$--$10$~keV
energy band. The CCD has the same design with that of the
EPIC MOS instruments of XMM-Newton. The size of the
field of view is $\timeform{23.'6} \times \timeform{23.'6}$.
In these observations, the XRT was operated with the Photon
Counting (PC) mode.  The observation logs are also
summarized in table~\ref{table:observation_log}.

We analyzed the cleaned event data which were generated by
the Swift Data Center (SDC) processing. We used the
\texttt{HEAsoft version 6.16} to analyze the Swift XRT
data.  Also, we used a relevant calibration database
(\texttt{CALDB Version 1.8}).

\begin{table*}
\begin{center}
\caption{Summary of the observations for GRS~1747$-$312}
\label{table:observation_log}
\begin{tabular}{lccccc}
\hline
Observatory & Obs.ID & Start time (UTC) & Exposure (ks)\footnotemark[$*$] & Outburst\footnotemark[$\dagger$]& Eclipse\footnotemark[$\ddagger$]\\
\hline
Chandra & 4551 & 2004/03/29 22:53:48 & $45.0$(ACIS-S/HETG) & $\surd$ & $\surd$ \\
XMM-Newton & 206990101 & 2004/09/28 13:18:58 & $14.2$ (MOS1/2), $8.2$ (pn) &  &  \\
Suzaku & 504092010 & 2009/09/16 07:21:35 & $45.3$ (XIS), $45.6$ (PIN) & & $\surd$ \\
Swift & 00032761001 & 2013/03/18 21:01:13 & $0.98$(XRT)& $\surd$ &  \\
Swift & 00032761002 & 2013/03/24 02:13:18 & $0.98$(XRT) & $\surd$ & $\surd$ \\

\hline
\multicolumn{4}{@{}l@{}}{\hbox to 0pt{\parbox{180mm} {\footnotesize
 \footnotemark[$*$] Effective exposure of the screened data; the HXD-PIN exposure is dead-time corrected.\\
 \footnotemark[$\dagger$] Observation during the outbursts.\\
 \footnotemark[$\ddagger$] Observation including the predicted time of eclipses. 
 }\hss}}
\end{tabular}
\end{center}
\end{table*}

\section{Analysis and results}
 \label{section:analysis}


\subsection{Image}
\label{subsection:image}

Although only GRS~1747$-$312 were detected
in Terzan 6 during the HRC observation,
there may be other transient X-ray sources 
in the cluster because of the high number density of the stellar objects.
Thus, in this section, 
we analyzed X-ray images of the observations
to investigate whether only GRS~1747$-$312 is visible or not.  Then we
defined source and background regions for the following
light curve and spectral analyses.

\subsubsection{Chandra}
\label{subsubsection:chandra_image}

\begin{figure}
 \begin{center}
 \FigureFile(80mm,50mm){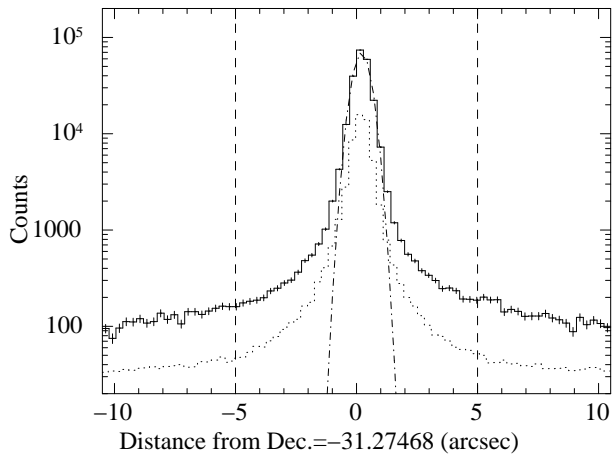}
  \caption{Projection profile along the declination axis
    obtained with the Chandra ACIS-S in the $0.5$--$10$~keV
    band. The bin size is $\timeform{0."28}$. The
    dash-dotted line shows the Gaussian function fitted to
    the profile between $\pm \timeform{1"}$.  The dotted line shows the profile simulated with \texttt{MARX}. To represent the background, a constant component is added. The normalization of the simulated profile is arbitrarily changed for the clarity.
    The region
    between the two dashed lines was defined as a source
    region.}
  \label{Chandra_Dec_pro}
 \end{center}
\end{figure}

Since the ACIS was operated in the CC mode, only a
one-dimensional image which is parallel to the long axis of
the ACIS-S can be obtained.  A peak of the brightness was
seen near the center of the image. Since the image
rotated only by $\timeform{2.4D}$ clockwise from the
north direction, we can study the position of the peak only along the
declination axis. A count profile along the declination axis
in the $0.5$--$10$~keV band binned with $\timeform{0."28}$
is shown in figure~\ref{Chandra_Dec_pro}.  The origin of the
horizontal axis is set to the position of GRS~1747$-$312,
$Dec. = -31.\!\!^{\circ}27468$, given by the Chandra HRC
observation \citep{2003A&A...406..233I}. We fitted the
profile between $\pm \timeform{1"}$ with a Gaussian function
and found that the peak position is $\delta Dec. =
+\timeform{0."208} \pm \timeform{0."001}$.  Since the 90\%
uncertainty circle of the Chandra X-ray absolute position
has a radius of
$\timeform{0."6}$~\citep{2003ExA....16....1W}, the peak
position is consistent with the position of GRS~1747$-$312.
The width of the Gaussian function was $\sigma =
\timeform{0."353 \pm 0."001}$. We simulated a profile of a
point source with
\texttt{MARX}~\citep{2012SPIE.8443E..1AD}.  The
width of a Gaussian function fitted to the simulated profile
was $\sigma = \timeform{0."358}$.  
A possibility of a spatially extended source or multiple objects
is thus rejected.
Though we could not study the image along the right
ascension axis, the observed source seems consistent with
GRS~1747$-$312.

In the following analysis, we extracted source photons from
$\timeform{-5"}$ to $\timeform{+5"}$ in the profile (see
figure~\ref{Chandra_Dec_pro}). Also, background events were
extracted from regions between $\timeform{-25"}$ and
$\timeform{-15"}$ and between $\timeform{+15"}$ and
$\timeform{+25"}$.

\subsubsection{XMM-Newton}

\begin{figure}
 \begin{center}
 \FigureFile(80mm,80mm){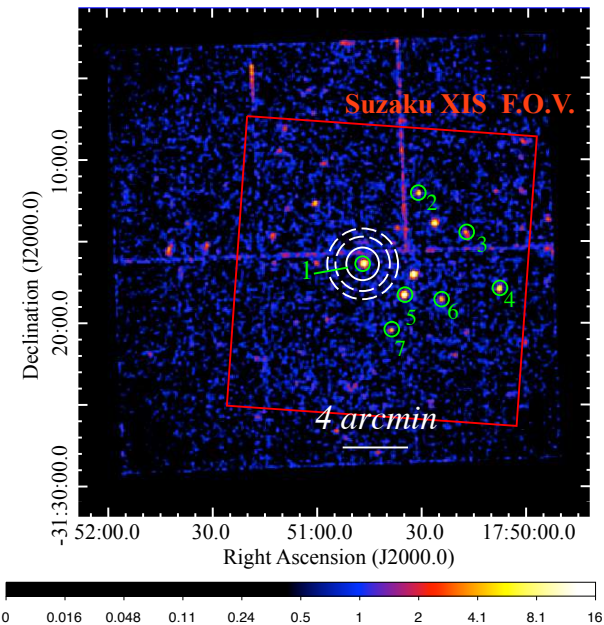}
  \caption{XMM-Newton pn image in the $0.5$--$10$~keV band
    smoothed with a Gaussian function of $\sigma =
    2$~pixels. The solid circle and dashed annulus indicate
    the source and background regions. The green circles
    show the sources listed in
    table~\ref{table:source_list}, which are used for 
the astrometry of the Suzaku image described in section~\ref{subsubsection:suzaku_image}. The red square
    corresponds to the field of view of the Suzaku XIS (see
    section~\ref{subsubsection:suzaku_image}).}
  \label{fig:xmm_image}
 \end{center}
\end{figure}

Figure~\ref{fig:xmm_image} shows an X-ray image in the
$0.5$--$10$~keV band obtained with the XMM-Newton pn. The
image was smoothed with a Gaussian function of $\sigma =
2$~pixels.

We found several sources in all pn, MOS1, and MOS2 images. Seven sources used later for the astrometry of the Suzaku image 
are indicated with green circles in figure~\ref{fig:xmm_image}. We designated them as Sources 1--7 hereafter.
These sources were tagged as point sources by the source detection tool in the \texttt{SAS}, \texttt{emldetect}.
Their positions are listed in
table~\ref{table:source_list}.
The statistical error
on the Source~1 position is $\timeform{0."7}$, and
Source~1 is coincident with GRS~1747$-$312 within an
attitude uncertainty of XMM-Newton ($3$--$\timeform{4"})$\footnote{XMM Users'
  Handbook:\\ http://www.mssl.ucl.ac.uk/www\_xmm/ukos/onlines/uhb\\ /XMM\_UHB/node104.html}.
Source~1 can thus be regarded as GRS~1747$-$312.

We defined a source region as a circle with a radius of
$\timeform{60"}$ centered on GRS~1747$-$312.  The source
region is shown as the white solid circle in
figure~\ref{fig:xmm_image}. A background region
was defined as its surrounding annulus with inner and outer radii of
$\timeform{100"}$ and $\timeform{130"}$. 
The center of the
annulus is the same as that of the circle of the source
region.

\subsubsection{Suzaku}
\label{subsubsection:suzaku_image}

X-ray images in the 0.5--10~keV band were obtained
with XIS0, 1, and 3. Since an X-ray source at the center of
the images exhibited a burst and a dip-like behavior in the
X-ray light curve as described in
section~\ref{subsection:light_curve}, we excluded the
durations of the burst and dip to create the XIS images.
The images of XIS0, 1, and 3 were combined, and binned with
$8 \times 8$ pixels.

\begin{figure}
 \begin{center}
 \FigureFile(80mm,80mm){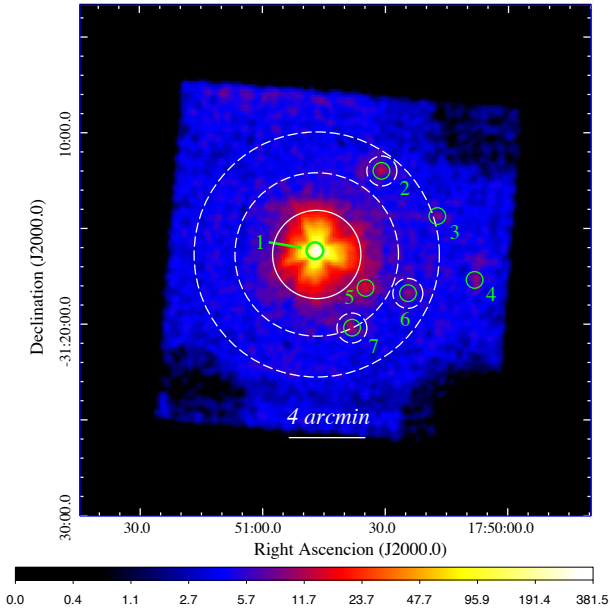}
  \caption{Combined image of three XIS sensors in the
    $0.5$--$10$~keV band after the coordinate correction
    as described in the text. X-ray events during the burst
    and dip were not used. Regions illuminated by the
    $^{55}$Fe calibration sources were removed.  The image
    was binned with $8 \times 8$ pixels and then smoothed
    with a Gaussian function of $\sigma = 2$~bins.  
    The color bar at the bottom of the image is in unit of
    photons per binned pixel. The solid circle and the dashed annulus indicate 
    the source and background regions. Green circles shows 
    the sources listed in table~\ref{table:source_list}.}
  \label{fig:suzaku_image}
 \end{center}
\end{figure}

All seven sources in table~\ref{table:source_list} can be
seen in the Suzaku image. We then checked the astrometry
between the Suzaku and XMM-Newton observations using these
sources. 
We made projection profiles of these sources along the right ascension and declination axes from the Suzaku image. These profiles were
fitted with a Gaussian function plus a constant component to determine the peak positions.
Since Sources 4 and 5 were too faint to fit with
the Gaussian function, we defined 
the pixel position having the largest count
as the position of these sources.
We found that the Suzaku positions were shifted on average
by {$\timeform{-1."1}$} along the right ascension
axis and by {$\timeform{+10."0}$} along the
declination axis.  We then corrected the coordinates of the
Suzaku image by these amounts. The Suzaku image thus
obtained is shown in figure~\ref{fig:suzaku_image}, and the
source positions are listed in
table~\ref{table:source_list}.

\begin{table*}[bhtp]
\begin{center}
\caption{List of sources detected with XMM-Newton and Suzaku.
}
\label{table:source_list}
\begin{tabular}{cccccc}
\hline
& \multicolumn{2}{c}{XMM-Newton image} &\multicolumn{2}{c}{Suzaku XIS image} & pn count rate \\
Number\footnotemark[$*$]  & RA & Dec & RA & Dec & (cts s$^{-1}$)\\
\hline
1 & $\timeform{267D.6950}$ & $\timeform{-31D.2750}$ & $\timeform{267D.697}$ & $\timeform{-31D.276}$ & $3.6 \pm 0.3$\\
2 & $\timeform{267D.6301}$ & $\timeform{-31D.2033}$ & $\timeform{267D.630}$ & $\timeform{-31D.205}$ & $1.6 \pm 0.2$ \\
3 & $\timeform{267D.5733}$ & $\timeform{-31D.2435}$ & $\timeform{267D.572}$ & $\timeform{-31D.244}$ & $1.1 \pm 0.2$ \\
4 & $\timeform{267D.5335}$ & $\timeform{-31D.3004}$ & $\timeform{267D.535}$ & $\timeform{-31D.301}$ & $2.7 \pm 0.3$\\
5 & $\timeform{267D.6468}$ & $\timeform{-31D.3071}$ & $\timeform{267D.646}$ & $\timeform{-31D.308}$ & $3.7 \pm 0.3$ \\
6 & $\timeform{267D.6026}$ & $\timeform{-31D.3114}$ & $\timeform{267D.603}$ & $\timeform{-31D.311}$ & $1.2 \pm 0.2$ \\
7 & $\timeform{267D.6625}$ & $\timeform{-31D.3430}$ & $\timeform{267D.661}$ & $\timeform{-31D.339}$  & $0.9 \pm 0.2$\\
 \hline
 \multicolumn{4}{@{}l@{}}{\hbox to 0pt{\parbox{180mm} {\footnotesize
 \footnotemark[$*$]  The numbers correspond to the ones in figures~\ref{fig:xmm_image} and \ref{fig:suzaku_image}.\\
 }\hss}}
\end{tabular}
\end{center}
\end{table*}

The Suzaku position of Source~1 is coincident with
GRS~1747$-$312 within the Suzaku attitude uncertainty of $\sim
{\timeform 19"}$~\citep{2008PASJ...60S..35U}. 
The radial profile of Source~1 was fitted with a point
spread function extracted from the Suzaku data of SS~Cyg.
The profile is consistent
with the point spread function
with $\chi^2/\nu = 1.12$ ($\nu = 294$), where $\nu$
represents the degree of freedom in the fitting.
Thus Source~1 can be regarded as GRS~1747$-$312.

Source 5 was located $\sim 3'$ away from GRS 1747$-$312 in the south-west direction, and we defined a source region as a circle
with a radius of $\timeform{2.'3}$ centered on
GRS~1747$-$312 to avoid the contamination from source~5.
The background events were extracted from an annulus with
inner and outer radii of $\timeform{4.'27}$ and
$\timeform{6.'45}$.  Since Sources 2, 6 and 7 were located
in the background region, we excluded X-ray photons within circles with a radius of $\timeform{0.'78}$, centered on these sources.
These regions are plotted as dashed
circles in figure~\ref{fig:suzaku_image}.

\subsubsection{Swift}

\begin{figure*}
 \begin{center}
 \FigureFile(80mm,80mm){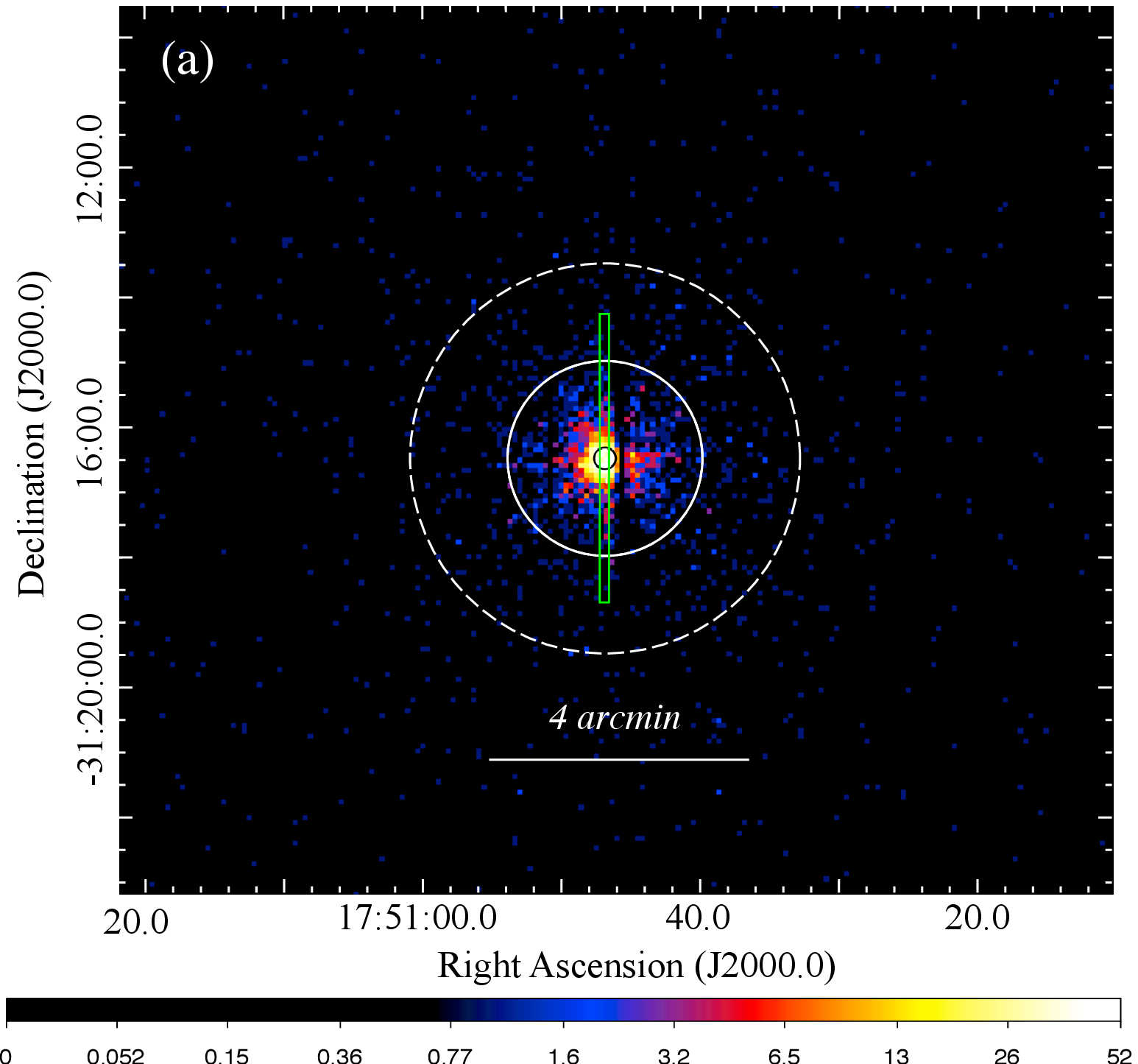}
 \FigureFile(80mm,80mm){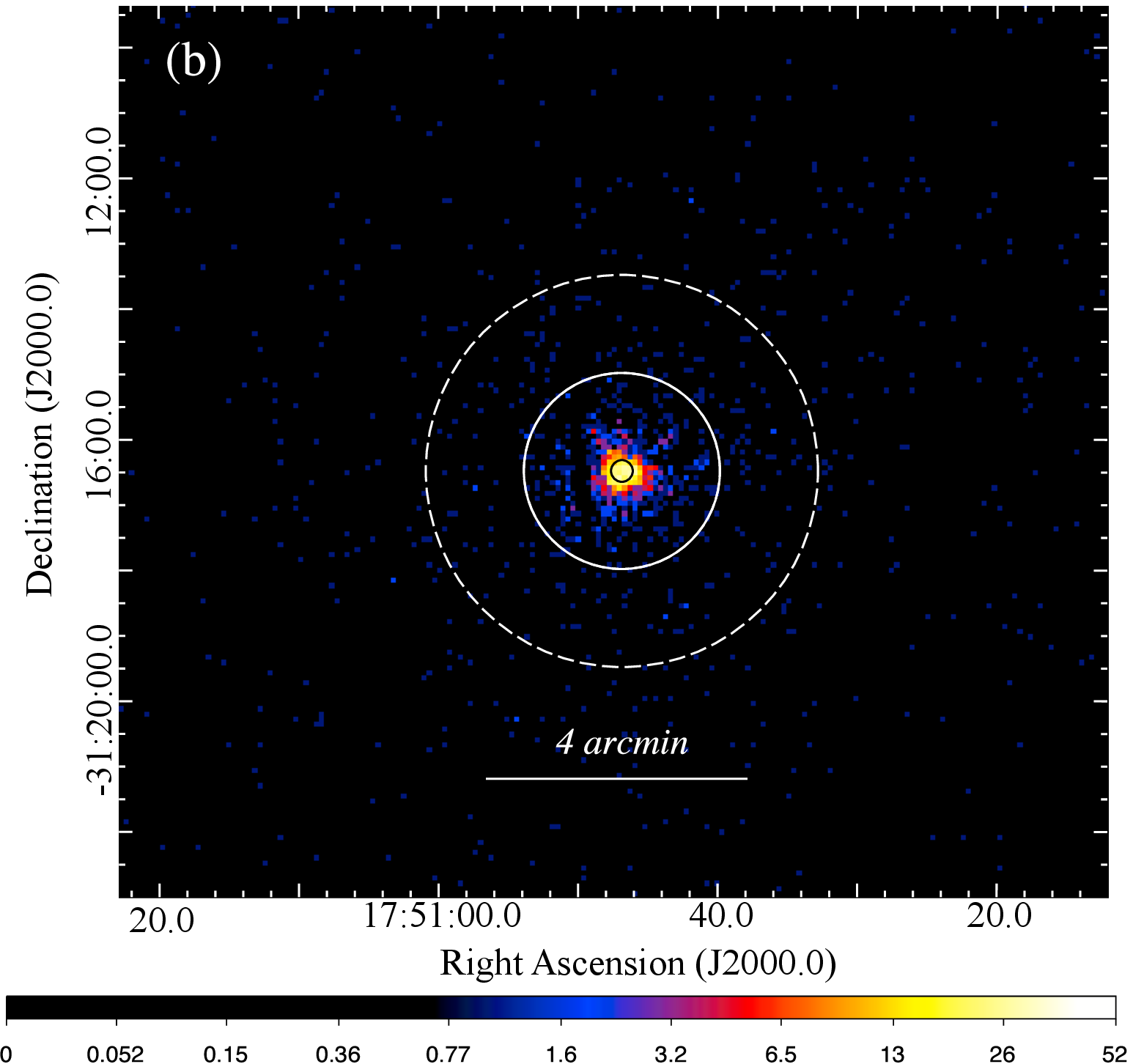} 
  \caption{Swift XRT images obtained during the 1st (a) and
    2nd (b) observations in the $0.5$--$10$~keV band. The
    region around GRS~1747$-$312 were zoomed up. The images
    were binned with $2 \times 2$~pixels. White solid and
    dashed circles show the source and background regions,
    respectively. We made the projection profile shown in
    figure~\ref{fig:swift_pro} from the green rectangle. The
    black circles show the regions in which events were
    removed from the source regions to mitigate the pile-up
    effect.}
  \label{fig:swift_image}
 \end{center}
\end{figure*}

Figures \ref{fig:swift_image}a and \ref{fig:swift_image}b show X-ray images
in the $0.5$--$10$~keV band obtained with the Swift XRT during
the 1st and 2nd observations, respectively. The images were binned
with $2\times 2$~pixels. 
Since there was a sudden decrease in the flux during the
2nd observation (see section~\ref{subsubsection:swift_lc}), we
filtered out the interval of the low-flux state to improve the signal-to-noise ratio. In both images, the brightness
peak was located near the center of the field of view.

\begin{figure}
 \begin{center}
 \FigureFile(80mm,50mm){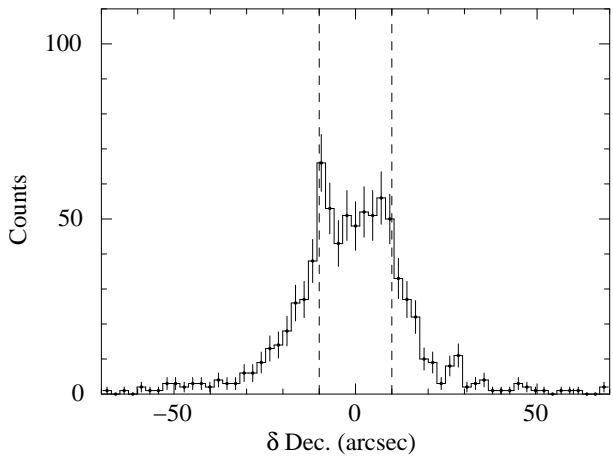}
  \caption{Projection profile during the Swift 1st observation
    extracted from the green rectangular in
    figure~\ref{fig:swift_image}~(a). The origin of the
    horizontal axis was set to be the position of
    GRS~1747$-$312. The bin size was $1$~pixel
    ($\timeform{2."36}$). 
X-ray events between the dashed lines 
were affected by the pileup, and they were
    not used in the following light curve and spectral
    analysis.}
  \label{fig:swift_pro}
 \end{center}
\end{figure}

To determine the position of the X-ray source, we made
projection profiles along the right ascension and
declination axes, and then fitted the profiles with a
Gaussian function. The projection profile of
the 1st observation along the declination axis obtained from
the green rectangular region in
figure~\ref{fig:swift_image}a is shown in
figure~\ref{fig:swift_pro}. 
The profile within $\pm \timeform{10''}$ looks suppressed. 
The suppression was also seen in the 2nd observation.
This is because of the pile-up effect, since the X-ray source was too bright 
($>2$~cts~s$^{-1}$).  
Although the profiles were not fully reproduced with a simple Gaussian
function, the Gaussian fittings gave the source position as
$(RA,Dec)_{\rm J2000.0} $ = ($267.\!\!^{\circ}6964$,
$-31.\!\!^{\circ}2741$) for the 1st observation and
($267.\!\!^{\circ}6955$, $-31.\!\!^{\circ}2748$) for the 2nd
observation.  These values are coincident with the position
of GRS~1747$-$312 within the fitting error ($\sim
\timeform{1"}$) and positional uncertainties of Swift ($\sim \timeform{3"})$\footnote{Documentation for Swift XRT
  GRB positions: http://www.swift.ac.uk/sper/docs.php}.
Hence, we consider the X-ray source detected with the Swift XRT as GRS 1747$-$312.

We found an X-ray source at almost the same position as GRS 1747$-$312 even during the low-flux state in the 2nd observation.
However, 
the statistics were too poor to do the above analysis. 
The pixel position with the maximum counts was $(RA,Dec)_{\rm J2000.0} = (267.\!\!^{\circ}6925, -31.\!\!^{\circ}2767)$.
There was no significant change of the position from that during the high-flux state.

For the light curves and spectral analyses described in the
following sections, we defined a source region as an annulus
with an outer radius of $\timeform{1.'5}$ centered on the position
of GRS~1747$-$312.  A circular region with a radius of
$\timeform{10"}$ was excluded from the source region to
remove the pileup events.
A background region was defined as an annulus with
inner and outer radii of $\timeform{1.'5}$ and
$\timeform{3.'0}$. The center of the annulus was the same as
that of the source region.  We made ARFs using \texttt{xrtmkarf} in \texttt{FTOOLS}, taking into account the
bad columns at ${\rm DETX} =290$--$294$ and
$319$--$321$ \footnote{Swift XRT
  thread:http://www.swift.ac.uk/analysis/xrt/digest\_cal.php}
  in the detector coordinate.

\subsection{Light curve}
\label{subsection:light_curve}

\subsubsection{Chandra}
\label{subsubsection:chandra_lc}

\begin{figure}
 \begin{center}
 \FigureFile(80mm,50mm){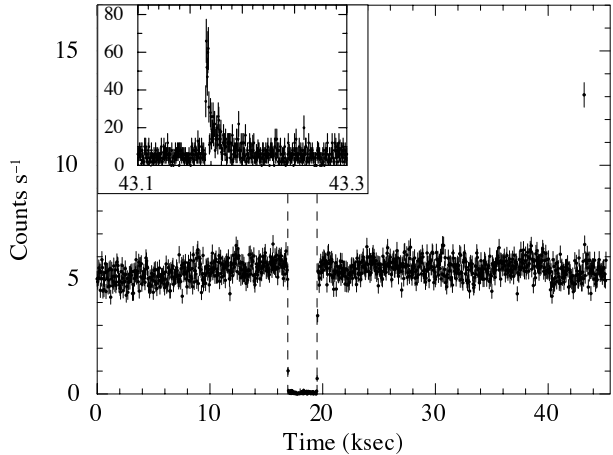}
  \caption{Light curve obtained with the Chandra observation
    in the $0.5$--$10$~keV band. The bin size is
    $46$~s. The time origin of the plot was MJD~$53093.96450$. 
    Dashed lines show the predicted time of the
    ingress and egress of the eclipse based on
    \citet{2003A&A...406..233I}. Top-left panel shows the
    zoomed up light curve around the
    burst.  \label{fig:chandra_lc}}
 \end{center}
\end{figure}

We extracted light curves in the $0.5$--$10$~keV band from
the source and background regions defined in
section~\ref{subsubsection:chandra_image}.  The background
light curve was subtracted from the source light curve after
normalizing the difference of the area.
Figure~\ref{fig:chandra_lc} shows the background-subtracted
light curve binned with $46$~s. The light curve is plotted
in Terrestrial Time corrected to solar system barycenter
(Barycentric Dynamical Time; TDB). The origin of the light
curve is MJD~$53093.96450$. In the light curve, a sudden
decline in the count rate from $17$~ks to
$19.5$~ks and a burst at $43$~ks were seen. The folding
time of the exponential decay of the burst was
$9.4^{+2.8}_{-2.2}$~s. The decay time scale is consistent
with that of a typical type-I X-ray burst.

In \citet{2003A&A...406..233I}, the ephemeris of
GRS~1747$-$312 was calculated as ${\rm MJD}
\,\,52066.259473(5) + 0.514980303(7)n$ for ingress and ${\rm
  MJD} \,\,52066.289497(10) + 0.514980303(7)n$ for egress
($n$ is an integer).
Based on the ephemeris, the ingress and egress of
eclipse were predicted at MJD~$53094.16016$ and
$53094.19020$ during the Chandra observation, respectively, which are shown with the dashed lines in
figure~\ref{fig:chandra_lc}.  Both of the observed ingress
and egress were coincident with the predicted time within
$\sim 10$~s.
Averaged count rates outside and during the eclipse were
$5.43 \pm 0.02$~cts~s$^{-1}$ and $0.05 \pm 0.01$~cts~s$^{-1}$, 
respectively. In this estimation,
time intervals of $30$~s
before and after the ingress and egress were
neglected to exclude the transition phases.

\subsubsection{XMM-Newton}

We made light curves of the pn, MOS1 and 2 from the source
and background regions in the $0.5$--$10$~keV band.  In the
background-subtracted light curves, no clear variability was
seen. The averaged count rate was $0.029 \pm 0.003$~cts~s$^{-1}$ 
for pn and $0.027 \pm 0.002$~cts~s$^{-1}$ for MOS $1+2$.

\subsubsection{Suzaku}
\label{subsubsection:suzaku_lc}

\begin{figure}
 \begin{center}
 \FigureFile(80mm,){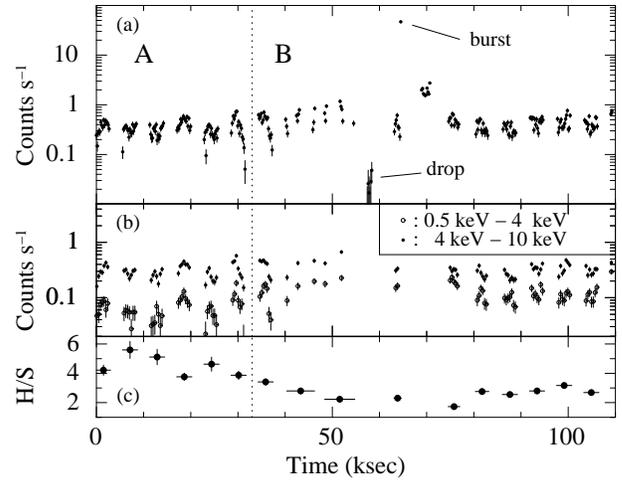}
  \caption{(a) Suzaku light curve of XIS0$+$3 
    in the $0.5$--$10$~keV band.  The bin size is $192$~s. 
    The time origin is MJD~$55090.30904$. 
    (b) Light curves of the XIS0$+$3 in
    the $0.5$--$4$~keV band (open circles) and $4$--$10$keV
    band (filled circles). The bin size is $384$~s. The
    intervals of the drop and burst were removed from the
    plot. (c) Hardness ratio of these energy bands. Each
    bin has approximately $1500$~photons in the
    $0.5$--$10$~keV band. We divided the exposure time into
    two intervals A and B with the dotted line.}
  \label{fig:suzaku_lc}
 \end{center}
\end{figure}

We made XIS0 $+$ XIS3 light curves from the source and
background regions in the $0.5$--$10$~keV band. Figure~\ref{fig:suzaku_lc}a
shows the background-subtracted light curve where the bin
size is $192$~s. The origin of the light curve is MJD~$55090.30904$ in TBD.
We can see fluctuations of the count rate in the light curve.
Also a remarkable drop and a burst were seen at around $\sim 57$~ks and $\sim
64$~ks, respectively.  The burst lasted for a few hours. 
The detailed analysis and discussion on the burst 
is reported in \citet{2014efxu.conf..158I}.

During the drop, the combined count rate of XIS$0 + 3$ was
$0.025 \pm 0.026$~cts~s$^{-1}$.
 We could not observe the start and end of the
drop due to the Earth occultation or the SAA passage. The
upper and lower limits for the duration of the drop were
$8296$~s and $1368$~s, respectively. 
Some dips are reported  in \citet{2003A&A...406..233I}. However, the duration of the dipping activity was $\sim 50$~s, which is quite shorter than that of the drop observed with Suzaku.
The averaged count rate excluding the burst and drop was $0.425 \pm 0.005$~cts~s$^{-1}$.

\begin{figure}
 \begin{center}
 \FigureFile(80mm,){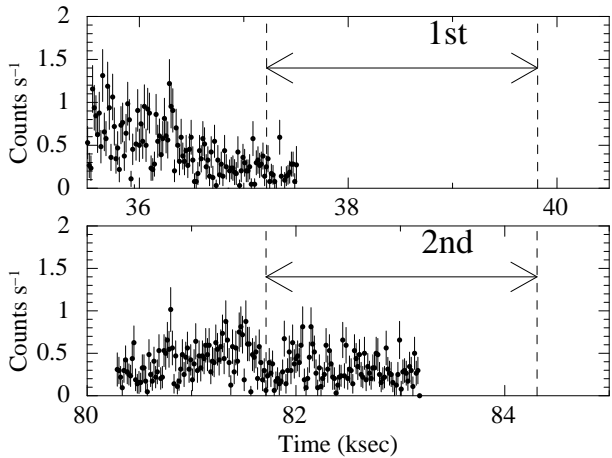}
  \caption{Zoomed-up light curves in the $0.5$--$10$keV band around the predicted
    eclipses during the Suzaku observation. The time origin is MJD~$55090.30802$ and the bin size is $16$~s. The dashed lines show the
    predicted time of the ingress and egress of the
    eclipses. The arrows indicate the intervals of the
    predicted eclipses.}
  \label{fig:suzaku_lc_zoom}
 \end{center}
\end{figure}

\begin{figure}
 \begin{center}
 \FigureFile(80mm,){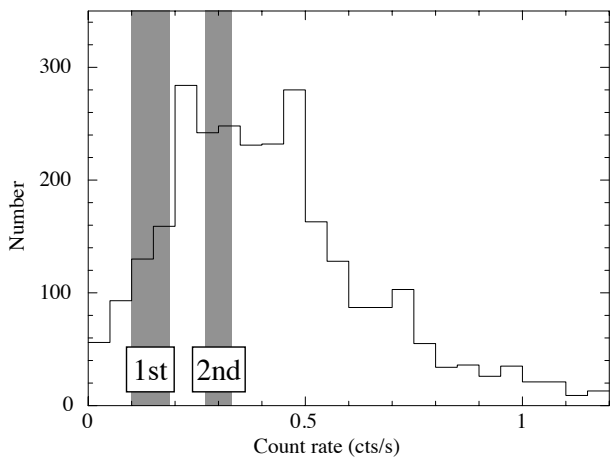}
  \caption{Histogram of the count rate in the $0.5$--$10$keV band during the Suzaku observation, excluding the burst and drop, and the duration of the predicted eclipses. 
To make the histogram, we used the count rates binned with $16$~s. The gray areas indicate the range of the count rates during the predicted eclipses.
}
  \label{fig:suzaku_histogram}
 \end{center}
\end{figure}

During the Suzaku observation, two eclipses were predicted,
from MJD~$55090.73879$ to $55090.76884$ and from
$55091.25377$ to $55091.28382$. The expected times of the
ingresses and egresses are shown with dashed lines in the zoomed
up light curve (figure~\ref{fig:suzaku_lc_zoom}).  Arrows in
the figure show the predicted intervals of the eclipses.
Due to both the Earth occultation and the SAA passage, only the first $288$~s and $1464$~s of the predicted intervals were observed. In contrast to the previous researches and 
our Chandra result in section~\ref{subsubsection:chandra_lc},
no large decrease in the count rate was seen during 
the both predicted eclipses.
The count rates were $0.14\pm0.04$~cts~s$^{-1}$ in the first predicted
eclipse and $0.30\pm0.03$~cts~s$^{-1}$ in the second one.
On the other hand, the count rate in an interval of $288$~s just before the first ingress was $0.25 \pm 0.06$~cts s$^{-1}$,  and that of $1464$~s just before the second ingress was $0.45 \pm 0.03$ cts s$^{-1}$. The count rates during the predicted eclipses are thus lower than the count rates before the eclipses.
The low count rates might suggest that weak eclipses were observed. However, the average duration of the transition for ingress was $27.3 \pm 0.8$~s, and the mid point of the transition occurred within $+1.8$~s/$-2.5$~s from the predicted time~\citep{2003A&A...406..233I}.
The prediction about the time of the transition mid point could be affected by the change of the orbital period and the uncertainty of the orbital period.  The upper limit of $\dot{P}/P$ was estimated to be $1 \times 10^{-16}$~s$^{-1}$ at MJD 52066.259477~\citep{2003A&A...406..233I}. GRS 1747-312 had completed  5873 revolutions since then, and the change of the orbital period could affect the predicted times of the mid points by 3.4 s at most. The uncertainty of the orbital preiod is $7 \times 10^{-9}$~d $= 6 \times 10^{-4}$~s~\citep{2003A&A...406..233I}.  After the 5873 revolutions, the accumulation of the uncertainty could change the predicted time of the mid points by 3.6s at most. Thus the mid point  of the transition should be coincident with the prediction within $+8.8$~s/$-9.5$~s.
The Suzaku light curve does not show these characteristics clearly. Note that the count rate  during the Suzaku observation was highly fluctuated. Figure~\ref{fig:suzaku_histogram} shows a histogram of the count rate excluding the burst, drop, and the durations of the predicted eclipses.
The count rate sometimes became lower than that during the interval of the eclipses. Thus the low count rates during the predicted eclipses can be explained by the fluctuation, though the weak eclipse cannot be fully rejected.


Figure~\ref{fig:suzaku_lc}b shows the background
subtracted light curves in the 0.5--4~keV and 4--10~keV
bands. The bin size is 384~s. Figure~\ref{fig:suzaku_lc}c
shows the hardness ratio given by $H/S$, where $H$ and $S$
represent the count rates in the $4$--$10$~keV and
$0.5$--$4$~keV band, respectively.  The bin size was
adjusted so that $\sim 1500$ photons in the $0.5$--$10$~keV
band were included in each bin.  During the first $\sim
30$~ks of the observation, the count rate in the
$0.5$--$4$~keV band was small with $< 0.1$~cts~s$^{-1}$.
The count rate then increased to $\gtrsim 0.1$~cts~s$^{-1}$.  
On the other hand, the count rate in the $4$--$10$~keV band was
relatively stable. As a result, the hardness ratio decreased.

We divided the observation time into two intervals,
designated as A and B, with the dotted line in
figure~\ref{fig:suzaku_lc}. The time of the drop and the burst was removed from interval B. During interval~A, the hardness
ratio generally exceeded 4, 
while the hardness ratio decreased 
and then kept constant at $\sim 3$ during interval~B.

\subsubsection{Swift}
\label{subsubsection:swift_lc}

\begin{figure*}
 \begin{center}
 \FigureFile(80mm,50mm){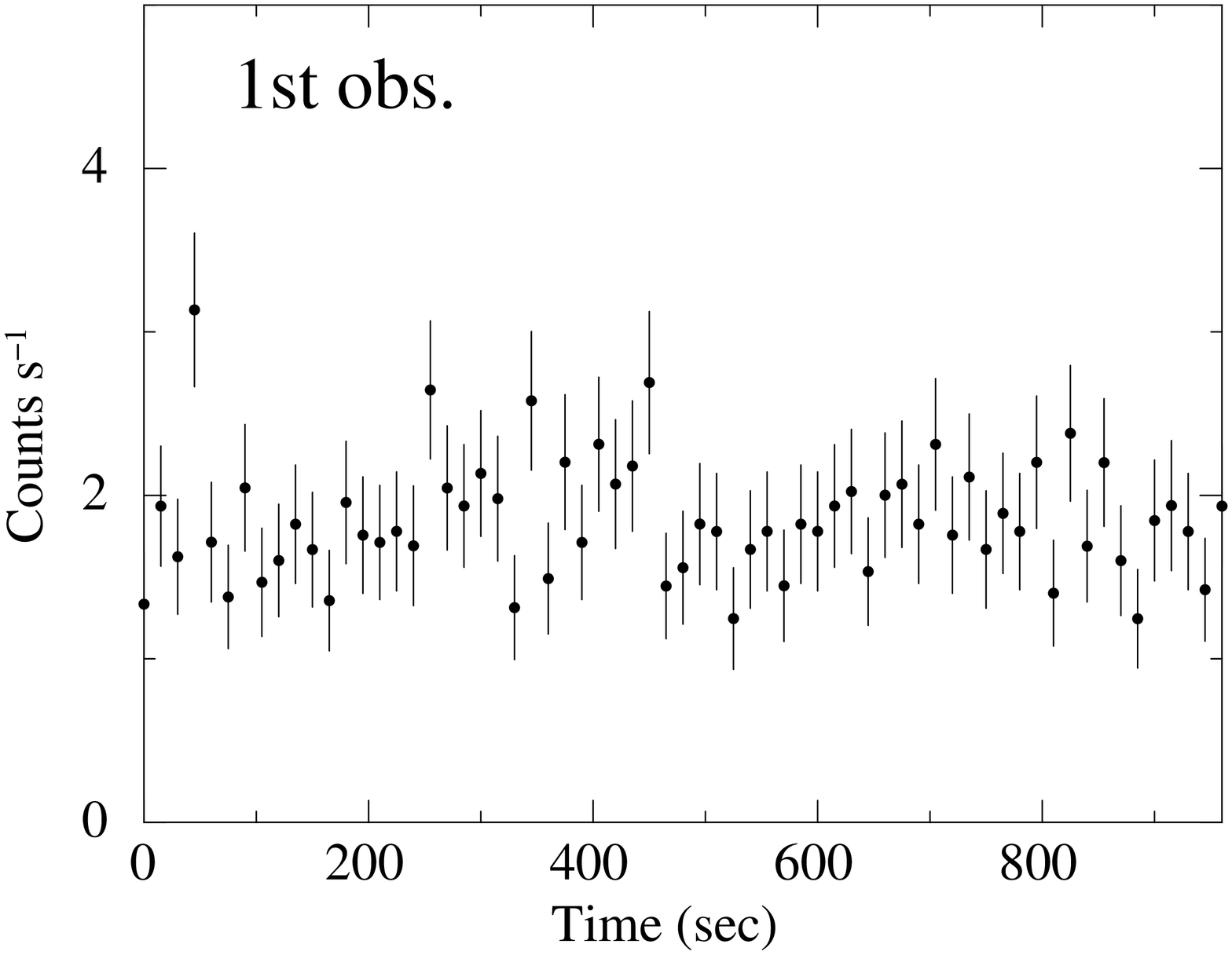}
 \FigureFile(80mm,50mm){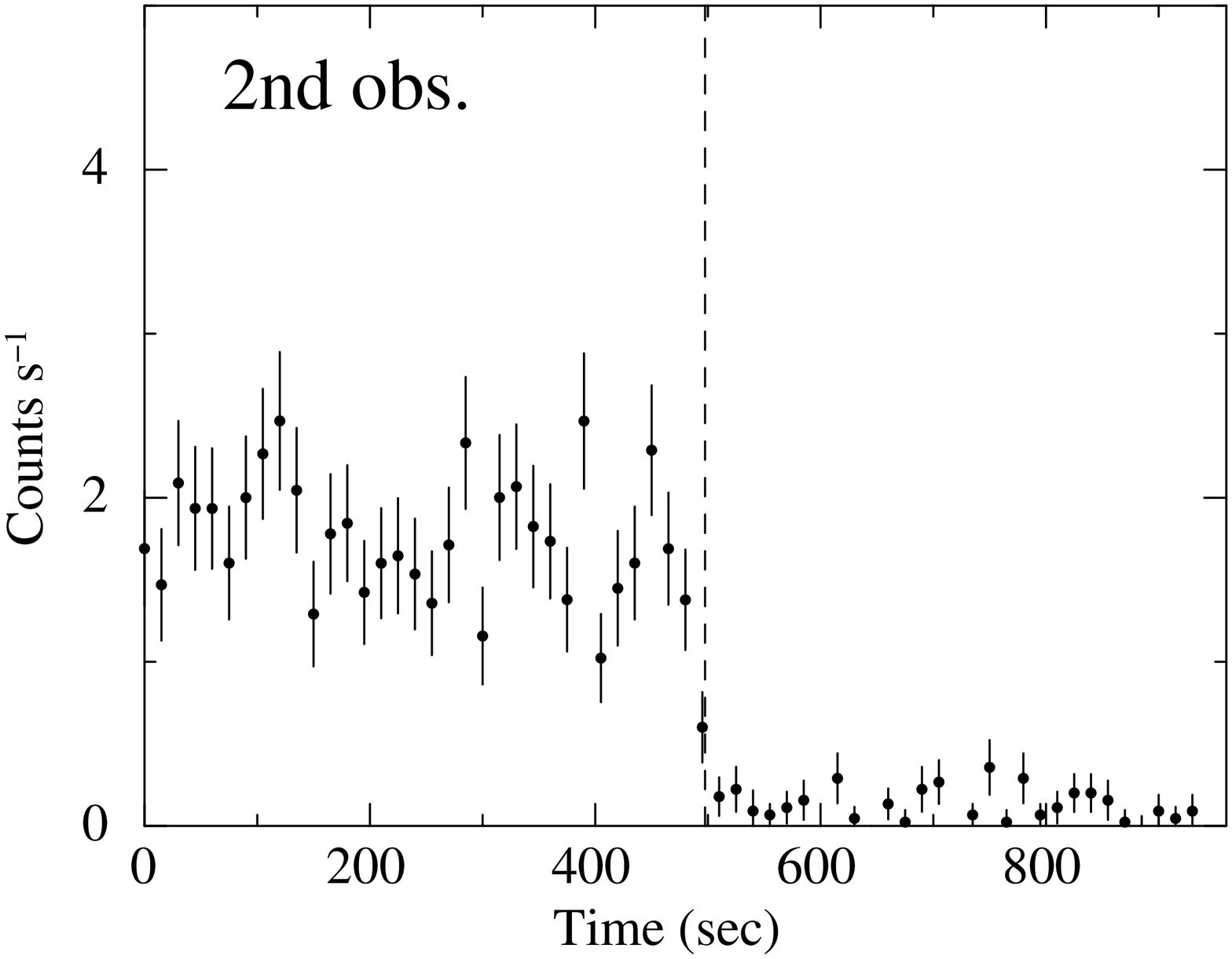}
  \caption{Swift light curves in the
    $0.5$--$10$~keV band during
    the 1st observation (left) and the 2nd observation (right). 
The bin size is $15$~s. The time
    origins are MJD~$56369.87667$ and $56375.09391$,
    respectively. The dashed line in the right panel shows
    the predicted time of the ingress of the
    eclipse.}
  \label{fig:swift_lc}
 \end{center}
\end{figure*}

Figure~\ref{fig:swift_lc} shows the background-subtracted
light curves in the $0.5$--$10$~keV band of each
observation. The curves are plotted in TDB and binned with
$15$~s. The time origins of these curves are MJD
$56369.87667$ and $56375.09391$, respectively.

During the 1st observation, the light curve had no remarkable
structure.  The averaged count rate was 
$1.85 \pm 0.07$~cts s$^{-1}$.  During the 2nd observation, 
on the other hand, a sudden decrease in the count rate was seen
at $\sim 500$~s.  The 2nd observation covered the time of the
predicted ingress of the eclipse at MJD~$56375.09967$
\citep{2003A&A...406..233I}, which is shown with the dashed
line in figure~\ref{fig:swift_lc}.  Thus the sudden decrease
was coincident with the expected time of the ingress.  
Averaged count rates 
before and after the decrease 
were $1.8 \pm 0.1$~cts~s$^{-1}$ and $0.10\pm0.03$~cts~s$^{-1}$,
respectively.

\subsection{Spectrum}
\label{subsection:Spectrum}

\subsubsection{Chandra}
\label{subsubsection:chandra_spec}

\begin{figure}
\begin{center}
 \FigureFile(80mm,){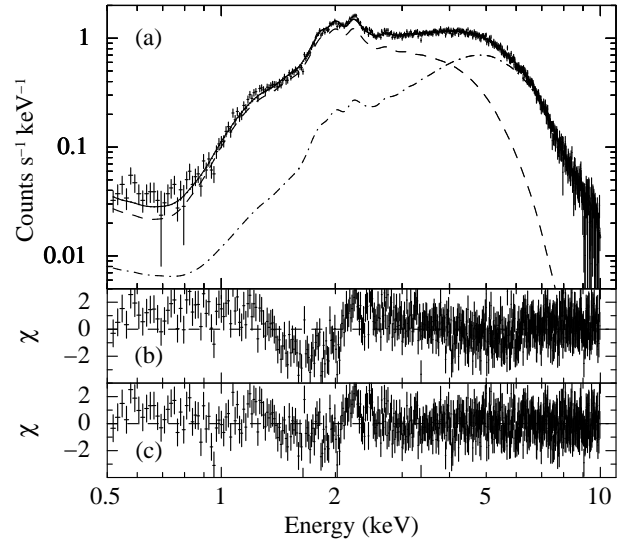}
  \caption{ (a) ACIS spectrum
    outside the eclipse with the best-fit blackbody (dashed) 
$+$ CompTT (dash-dotted) model. 
    (b) Residuals between the data and the best-fit single
    power-law model. (c) Residuals between the data and
    the blackbody $+$ CompTT model.
  \label{fig:chandra_spec}}
\end{center}
\end{figure}

We investigated the spectrum outside the eclipse
to know the properties during the outburst.  We extracted
the ACIS spectra from the source and background regions.
Figure~\ref{fig:chandra_spec} (a) 
shows the background subtracted spectrum in the $0.5$--$10$~keV band. 
We binned the spectrum so as to contain at least 10 photons in each
bin.

In the spectrum, no clear line-like features were seen. Then
we tried to describe the spectrum with an absorbed power-law
model, where the "phabs" model in the spectral fitting
package \texttt{XSPEC} \citep{1996ASPC..101...17A} was used for the
absorption model. This model adopts the cross sections of
\citet{1992ApJ...400..699B} with the solar abundance of
\citet{1989GeCoA..53..197A}. The result of the fitting was
not acceptable with $\chi^2/\nu = 1.49$ ($\nu =
634$). Figure~\ref{fig:chandra_spec} (b) shows the residuals
between the data and the best-fitted single power-law
model. There are large residuals below $\sim 3$~keV.
Neither a blackbody model ($\chi^2/\nu = 8.26$) nor a disk
blackbody model~(\cite{1984PASJ...36..741M}; \cite{1986ApJ...308..635M}) ($\chi^2/\nu = 2.47$) can describe the spectrum.

Then we examined several two-components models which are usually used to represent spectra of LMXBs.
First, we attempted to fit the data with a model which consists of a disk blackbody component  and a blackbody component from the surface of the neutron star.
The result of the fitting was not acceptable with $\chi^2/\nu = 1.25$ ($\nu = 632$). Second, we tried a model consisting of a disk blackbody and a Comptonized emission, assuming that only the photons from the surface of the neutron star are Comptonized. We used the "CompTT" model in
\texttt{XSPEC} \citep{1994ApJ...434..570T} to describe the Comptonized emission. In this Comptonization model, a spherical distribution was assumed for the Comptonizing hot plasma. 
This model was not rejected with  $\chi^2/\nu = 1.15$ ($\nu = 630$). The best-fit parameters were the following: the absorption column density $N_{\rm H} = 1.13\pm 0.06 \times 10^{-22}$~cm$^{-2}$, the temperature at the inner disk radius $kT_{\rm in} = 1.0^{+0.2}_{-0.1}$~keV, the temperature of the seed photons for the Comptonization $kT_{\rm seed} = 0.61^{+0.04}_{-0.03}$~keV, the temperature of the electrons for the Comptonization $kT_e = 8^{+7}_{-2}$~keV, and the plasma optical depth $\tau = 12\pm 2$.
Then we tested a "blackbody $+$ CompTT" model, in which  only the photons from the accretion disk are Comptonized. Since the electron temperature for the Comptonization was not sensitive in the
fitting, we fixed the value to that in the previous work, $kT_{\rm e} = 5.4$~keV \citep{2000A&A...355..145I}. This model was also not rejected with $\chi^2/\nu = 1.11$ ($\nu = 631$). The best-fit parameters are summarized in table~\ref{table:chandra_spec}.
We also checked the model that both of the surface of the neutron star and the accretion disk supply seed photons to the same Comptonizing hot plasma. Since this model was also not sensitive to the electron temperature, we fixed $kT_{\rm e}$ to $5.4$~keV. The model was not rejected with $\chi^2/\nu = 1.18$ ($\nu = 630$). We obtained the best-fit parameters as $N_{\rm H} = 0.94\pm 0.03 \times 10^{-22}$~cm$^{-2}$, $kT_{\rm seed}{\rm(disk)} = 0.69^{+0.02}_{-0.03}$~keV, $kT_{\rm seed}{\rm(neutron\,\,\,star)} = 2.28^{+0.09}_{-0.06}$~keV and $\tau < 1.2$.
Thus the last three models cannot be distinguished in terms of the statistics.
Since the "blackbody $+$ CompTT" model was applied for the
GRS 1747$-$312 spectrum obtained with BeppoSAX in the $0.5$--$100$~keV band in
\citet{2000A&A...355..145I}, we mainly investigate the model hereafter. 
Figures~\ref{fig:chandra_spec} (a) and (c)
show the best-fit model and the residuals 
between the data and model, respectively. The
absorption column density of $N_{\rm H} = (0.99 \pm 0.03)
\times 10^{22}$~cm$^{-2}$ was slightly lower than that for
Terzan 6, $N_{\rm H} = (1.2 \pm 0.1) \times 10^{22}$~cm$^{-2}$, which
was estimated via interstellar reddening
(\cite{1997A&AS..122..483B}; \cite{1995A&A...293..889P}).

\begin{table}[bhtp]
\begin{center}
\caption{Best-fit parameters for the Chandra ACIS spectrum.\label{table:chandra_spec}}
\begin{tabular}{lcc}
\hline
Interval & Outside eclipse & During eclipse \\
\hline
$N_{\rm H}$\footnotemark[$*$] & $0.99^{+0.03}_{-0.03}$ & $0.99$ (fixed)\\ \hline
$kT_{\rm BB}$~(keV) & $0.76^{+0.04}_{-0.06}$ & - \\
$R_{\rm BB}$~(km)\footnotemark[$\S$] & $8.1^{+0.4}_{-0.9}$ & - \\
$F_{\rm BB}$\footnotemark[$\dag$] & $1.82 \times 10^{-10}$ & - \\ \hline
$kT_{\rm seed}$~(keV) & $1.8^{+1.1}_{-0.7}$ & $1.8^{+0.5}_{-0.2}$ \\
$kT_{\rm e}$~(keV) & $5.4$~(fixed) & $5.4$~(fixed) \\
$\tau$ & $< 15$ & $16^{+8}_{-5}$ \\
$F_{\rm CompTT}$\footnotemark[$\dag$] & $3.87 \times 10^{-10}$ & $5.5 \times 10^{-12}$ \\ \hline
$\chi^2/\nu$ ($\nu$) & $1.11$ (631) & $0.78$ (15)\\
\hline
\multicolumn{3}{@{}l@{}}{\hbox to 0pt{\parbox{180mm} {\footnotesize
 \footnotemark[$*$] Hydrogen column density in units of 10$^{22}$~cm$^{-2} $.\\
 \footnotemark[$\S$] Radius of the blackbody sphere.\\
 \footnotemark[$\dag$] Observed flux in the $0.5$--$10$~keV band in the unit of erg cm$^{-2}$ s$^{-1}$.
  }\hss}}
\end{tabular}
\end{center}
\end{table}

\begin{figure}
\begin{center}
 \FigureFile(80mm,60mm){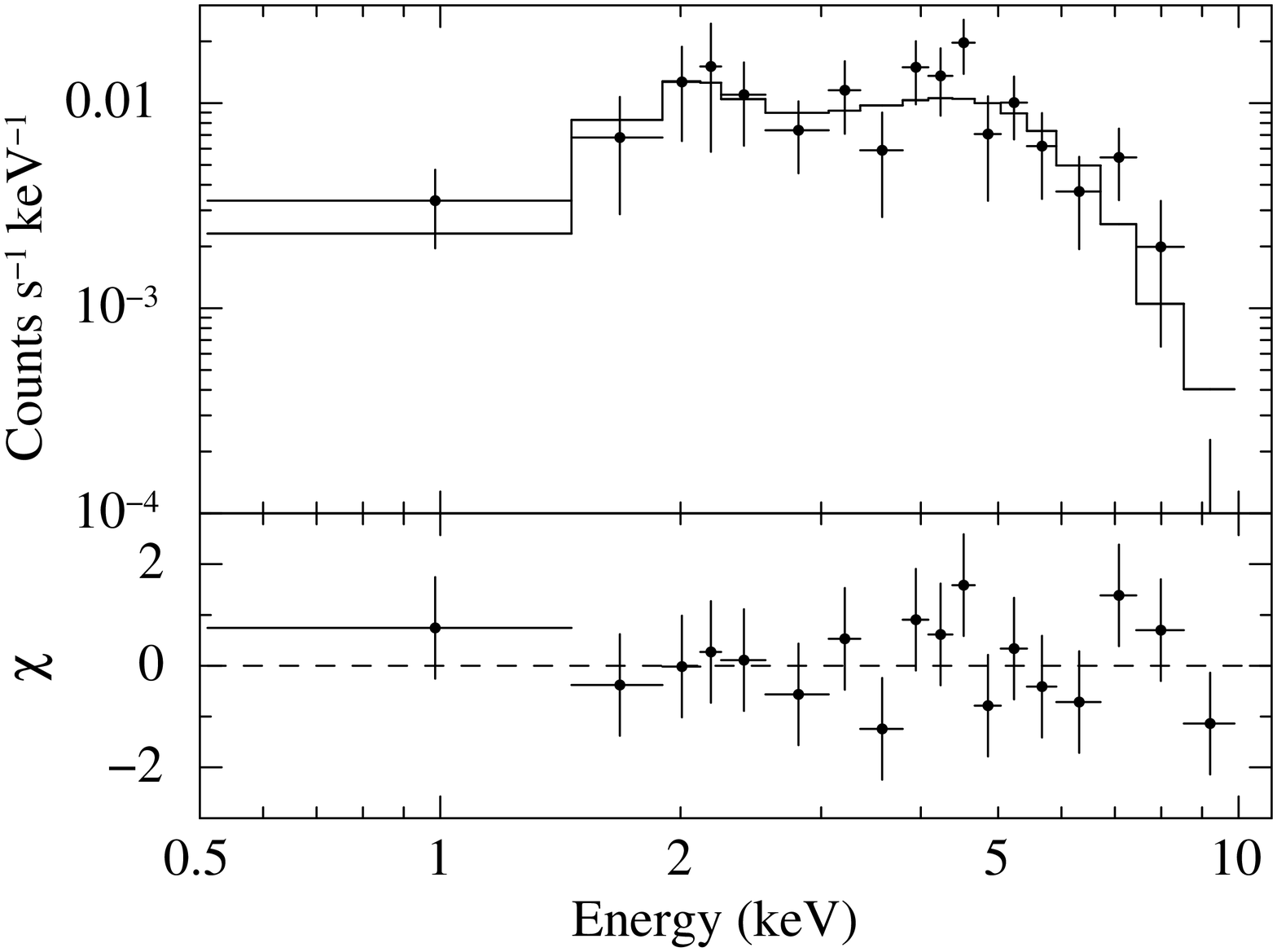}
  \caption{ ACIS-S spectrum during the eclipse
    fitted with an absorbed CompTT model in
    the $0.5$--$10$~keV band. Residuals
    between the data and the best-fit
    model are shown in the lower panel. 
\label{fig:chandra_eclipse_spec}}
\end{center}
\end{figure}

Next we studied a spectrum during the eclipse
(figure~\ref{fig:chandra_eclipse_spec}).  The spectrum was
binned so that each bin has at least $10$~photons.  We tried
to fit the spectrum with an absorbed CompTT
model. Similar to the fitting for the spectrum outside the
eclipse, the electron temperature of the CompTT model 
was fixed to $5.4$~keV.
The absorption was also fixed to the value outside the
eclipse, $N_{\rm H} = 0.99 \times 10^{22}$~cm$^{-2}$.  The
fitting was acceptable with $\chi^2/\nu = 0.78$ ($\nu = 15$).  
The best-fit model and residuals are shown in
figure~\ref{fig:chandra_eclipse_spec}, and the best-fit parameters
are summarized in table~\ref{table:chandra_spec}. 
There were no significant changes in the CompTT parameters, except for the normalization, from those obtained with the spectrum of the outside of the eclipse. However, due to its low photon
statistics, single component models such as a blackbody or
power-law could also describe the spectrum.  The flux
declined by a factor of $(1.0 \pm 0.2) \times 10^{-2}$
compared to the flux outside the eclipse.

\subsubsection{XMM-Newton}

\begin{figure}
\begin{center}
 \FigureFile(80mm,50mm){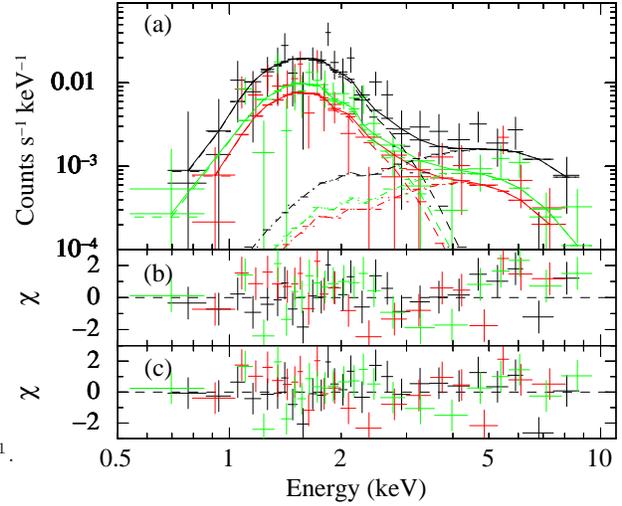}
  \caption{(a) XMM-Newton spectra obtained with pn (black),
    MOS1 (red) and MOS2 (green), with the best-fit
    blackbody (dashed) $+$ Comptonization (dash-dotted) model.  
(b) Residuals
    between the data and the best-fit power-law model. 
(c) Residuals between the data and the best-fit blackbody $+$ Comptonization model.
  \label{fig:xmm_spec}}
\end{center}
\end{figure}

Figure~\ref{fig:xmm_spec}(a) shows the 
spectra in the $0.5$--$10$~keV band. The black, red and
green data points show the spectra obtained with pn, MOS1 and
MOS2, respectively. The bin size of the spectra was set so
that each bin contains at least $15$ photons.

We tried to fit the spectra with an absorbed power-law
model. However, the fit was not acceptable with $\chi^2/\nu
= 1.57$ ($\nu = 67$) due to large residuals above
$\sim\ 2$~keV as shown in figure~\ref{fig:xmm_spec}~(b).
We then tried the blackbody plus CompTT model as in the case
of the analysis of the Chandra spectrum.  The electron
temperature of the CompTT model was fixed to 
$5.4$~keV. Since the temperature of the seed photons was
also insensitive in the fitting, we fixed the value to
$1.8$~keV, which is obtained in the analysis of the Chandra
spectrum.  The fitting was acceptable with $\chi^2/\nu =
1.38$ ($\nu = 64$), and the best-fit parameters are
summarized in table~\ref{table:xmm_spec}.

\begin{table}[bhtp]
\begin{center}
\caption{Best-fit parameters for the XMM-Newton spectra. \label{table:xmm_spec}}
\begin{tabular}{lc}
\hline
 Parameter & Best-fit value\\
\hline
$N_{\rm H}$\footnotemark[$*$] & $1.4^{+0.9}_{-0.6}$\\ \hline
$kT_{\rm BB}$~(keV) & $0.33^{+0.09}_{-0.07}$\\
$R_{\rm BB}$~(km)\footnotemark[$\S$] & $1.4^{+1.1}_{-0.4}$\\
$F_{\rm BB}$\footnotemark[$\dag$] & $6.7 \times 10^{-14}$ \\ \hline
$kT_{\rm seed}$~(keV) & $1.8$~(fixed)\\
$kT_{\rm e}$~(keV) & $5.4$~(fixed)\\
$\tau$ & $< 38$\\
$F_{\rm CompTT}$\footnotemark[$\dag$] & $1.45 \times 10^{-13}$ \\ \hline
$\chi^2/\nu$ ($\nu$) & $1.37$ (65)\\
\hline
\end{tabular}
\end{center}
\scriptsize
 \footnotemark[$*$] Hydrogen column density in units of $10^{22}$~cm$^{-2}$.\\
 \footnotemark[$\S$] Radius of the blackbody sphere.\\
 \footnotemark[$\dag$] Observed flux in the $0.5$--$10$~keV band in the unit of erg cm$^{-2}$ s$^{-1}$.
\end{table}

\subsubsection{Suzaku}
\label{subsection:suzaku_spec}

\begin{figure}
\begin{center}
 \FigureFile(80mm,50mm){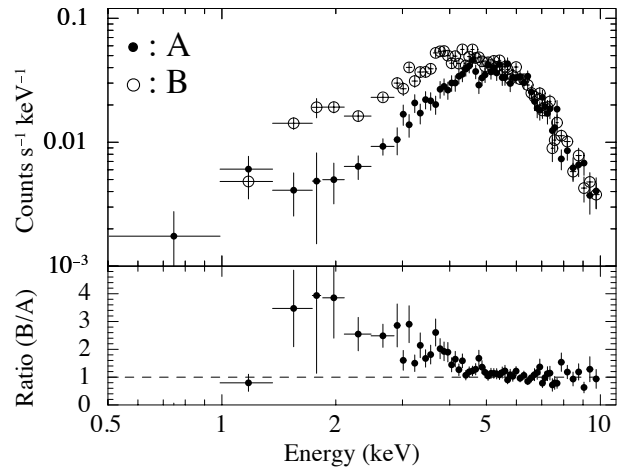}
  \caption{(Upper panel) Suzaku XIS1 spectra
    of intervals A (closed circles) and B (open
    circles). (Lower panel) Ratio between the spectra of
    intervals A and B.
  \label{spec_comp}}
\end{center}
\end{figure}

XIS spectra of intervals A and B were extracted from the
source and background regions, and the spectra were binned
so that each bin has at least 20 photons. The XIS1 spectra
of intervals A and B after the background subtraction are
shown in the upper panel of figure~\ref{spec_comp},
while the lower panel shows the ratio between the both spectra.  
The X-ray count rates below $4$~keV of interval B was found to be
higher than that of interval A .

\begin{table*}[bhtp]
\begin{center}
\caption{Best-fit parameters of the blackbody $+$ CompTT model
fitted to the Suzaku spectra.
\label{table:fit_result}
}
\begin{tabular}{lcccc}
\hline
 & \multicolumn{2}{c}{Independent fit} & \multicolumn{2}{c}{Simultaneous fit}\\ \hline
Interval & A & B & A & B \\
\hline
$N_{\rm H,BB}$\footnotemark[$*$] & $1.2$ (fixed) & $1.7^{+1.0}_{-0.9} $ & $1.2$ (fixed) & $1.8^{+0.5}_{-0.4}$\\
$kT$~(keV) & $0.4^{+0.3}_{-0.1}$ & $0.5^{+1.6}_{-0.2}$ & $0.4^{+0.1}_{-0.1}$ & $1.1^{+0.1}_{-0.1}$ \\
$R_{\rm BB}$~(km)\footnotemark[$\S$] & $1.2^{+0.4}_{-0.2}$ & $1.2^{+0.6}_{-0.2}$ & $1.3^{+0.3}_{-0.2}$ & $0.40^{+0.02}_{-0.02}$\\
$F_{\rm BB}$\footnotemark[$\dag$] & $8.1 \times 10^{-14}$ & $3.57 \times 10^{-13}$ & $7.86 \times 10^{-14}$ & $1.86 \times 10^{-12}$ \\ \hline
$N_{\rm H,CompTT}$\footnotemark[$*$] & $10^{+4}_{-2}$ & $12^{+2}_{-2}$&  \multicolumn{2}{c}{$11^{+3}_{-2}$} \\
$kT_{\rm seed}$ (keV) & $< 1.9$ & $< 1.4$ & \multicolumn{2}{c}{$1.1^{+0.3}_{-0.3}$} \\
$kT_{\rm e}$ (keV) & $5.4$~(fixed) & $5.4$~(fixed) & \multicolumn{2}{c}{$5.4$~(fixed)} \\
$\tau$ & $11^{+3}_{-6}$ & $10.9^{+0.7}_{-0.6}$ & \multicolumn{2}{c}{$12^{+2}_{-2}$} \\
$F_{\rm CompTT}$\footnotemark[$\dag$] & $0.99 \times 10^{-11}$ & $1.11\times10^{-11}$ & \multicolumn{2}{c}{$0.97\times 10^{-11}$} \\ \hline
$\chi^2/\nu$ ($\nu$) & $0.91$ (185) & $1.13$ (240) &  \multicolumn{2}{c}{$1.06$ (429)}\\
\hline
\multicolumn{5}{@{}l@{}}{\hbox to 0pt{\parbox{180mm} {\footnotesize
\footnotemark[$*$] Hydrogen column density in units of 10$^{22}$~cm$^{-2} $.\\
\footnotemark[$\S$] Radius of the blackbody sphere.\\
\footnotemark[$\dag$] Observed flux in the $0.5$--$10$~keV band in units of erg cm$^{-2}$ s$^{-1}$. The fluxes of the XIS0 1 and 3 were averaged.
  }\hss}}
\end{tabular}
\end{center}
\end{table*}

\begin{figure}
\begin{center}
 \FigureFile(80mm,50mm){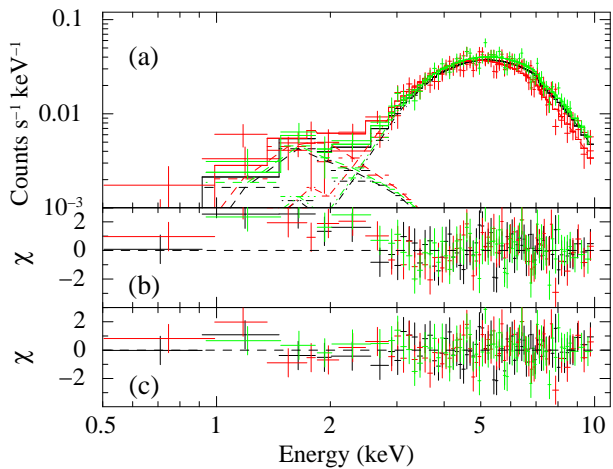}
  \caption{(a) Suzaku spectra of interval A.
The background-subtracted spectra obtained with XIS0, 1, and 3 
are represented by black, red and green, respectively.
The best-fit blackbody (dashed) $+$ CompTT (dash-dotted) model is also shown,
where the column densities for the two components are different
(see the text).
(b) Residuals between the data and  the best-fit single CompTT model.  
(c) Residuals between the data and the blackbody $+$ CompTT model.
  \label{A_fit}
}
\end{center}
\end{figure}

First, the spectrum of interval A was analyzed.  
The spectra
of XIS0, 1 and 3 were simultaneously fitted with an absorbed
CompTT model. Here we introduced constant factors, by which the normalization of each sensor was multiplied, to mitigate the
uncertainties of the XIS calibration.  The fit was not
acceptable with $\chi^2/\nu = 1.26$ ($\nu = 187$).
Significant residuals below 2~keV can be seen in
figure~\ref{A_fit} (b). 
The existence of a soft component is
shown by the Suzaku spectra as well as the Chandra and
XMM-Newton spectra.

Then we tried to fit the Suzaku spectra with the blackbody
$+$ CompTT model. We assumed that the electron temperature
of the Comptonization is 5.4~keV as in the case of the
Chandra and XMM-Newton spectra.  If we assumed the same
absorption for the two components, the fit was acceptable
with $\chi^2/\nu = 0.98$ ($\nu = 185$).
The best-fit parameters of the blackbody component
were $kT = 0.07^{+0.02}_{-0.01}$~keV,
$N_{\rm H} = 7.1^{+1.4}_{-1.1} \times 10^{22}$~cm$^{-2}$.
However, the lower limit of the bolometric luminosity
of the blackbody component was $1.0\times 10^{39}$~erg~s$^{-1}$,
which is significantly larger than the Eddington luminosity for a
$1.4 M_{\odot}$ neutron star, $\sim 3 \times 10^{38}$~erg~s$^{-1}$.

The model was then modified so that each component suffers
different absorption, given by "phabs1 $\times$ blackbody
$+$ phabs2 $\times$ CompTT".  We fixed the column density
for the blackbody component to the Galactic value 
$N_{\rm H, BB} = 1.2 \times 10^{22}$~cm$^{-2}$, since if we set the
absorption to be free, only an upper limit was obtained. 
This model can describe the spectrum
with $\chi^2/\nu = 0.91$ ($\nu = 185$).
The results of the fitting are summarized in
figures~\ref{A_fit} (a) and (c) and table~\ref{table:fit_result}.

\begin{figure}
\begin{center}
 \FigureFile(80mm,50mm){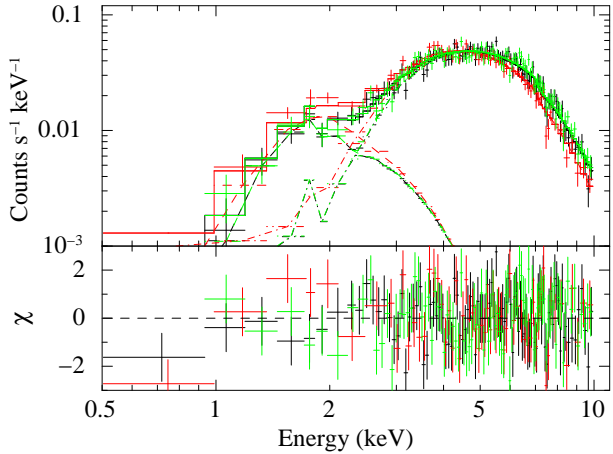}
  \caption{Suzaku spectra of interval B.
The best-fit blackbody $+$ CompTT model is also shown.
The definitions of the colors and lines are the
    same as those in figure~\ref{A_fit}.  
Residuals between the data and the best-fit model are shown in the lower panel.  }
  \label{B_fit}
\end{center}
\end{figure}

Secondly, the spectrum of interval B was analyzed. The
spectrum also could not be fitted with a single-component
model such as a blackbody or a CompTT model.  Thus we tried
the absorbed blackbody $+$ CompTT model where the column densities for
the two components were different as in the case of
interval~A.
The spectrum of interval B was well
described with this model with $\chi^2/\nu = 1.13$~($\nu = 241$).  
The results of the fitting are
summarized in figure~\ref{B_fit} and table~\ref{table:fit_result}.
Similar to interval A, the CompTT component
was strongly absorbed with a column density of $\sim 10^{23}$~cm$^{-2}$.

Finally, we simultaneously fitted the spectra of intervals
A and B with the blackbody $+$ CompTT model.
The fact that both spectra are quite similar to
each other above $\sim\ 5$~keV (figure~\ref{spec_comp})
suggests that both spectrum can be explained by the same
CompTT model, but they require a different blackbody component.
In this case, the fit was acceptable with $\chi^2/\nu = 1.06$~($\nu
= 420$).  The best-fit parameters are summarized in
table~\ref{table:fit_result}.

The absorption column density of $N_{\rm H} \sim 10^{23}$~cm$^{-2}$ was much higher than the value obtained from the spectrum during the Type-I burst, $1.46 \times 10^{22}$~cm$^{-2}$~\citep{2014efxu.conf..158I}. Thus we investigated the lower limits of the column density for several spectral models. When we fitted the spectrum of interval A with the "blackbody + CompTT" model with a free $kT_e$ value, a lower limit of the absorption column density of $7.6 \times 10^{22}$~ cm$^{-2}$ was obtained when $kT_{\rm e}$ was $2.4$~keV. We also tested the spectrum model of "blackbody + blackbody", then we obtained the lower limit of $N_{\rm H} = 7.8 \times 10^{22}$~ cm$^{-2}$. Thus we concluded that the absorption column density during the Suzaku observation was significantly higher than the interstellar absorption, independent on the spectral fitting models.

We also investigated the HXD PIN spectrum, but the spectrum
turned out to be significantly contaminated by another bright
source as described in appendix. Thus the HXD data are not used
in this paper.

\subsubsection{Swift}

\begin{figure}
\begin{center}
 \FigureFile(80mm,50mm){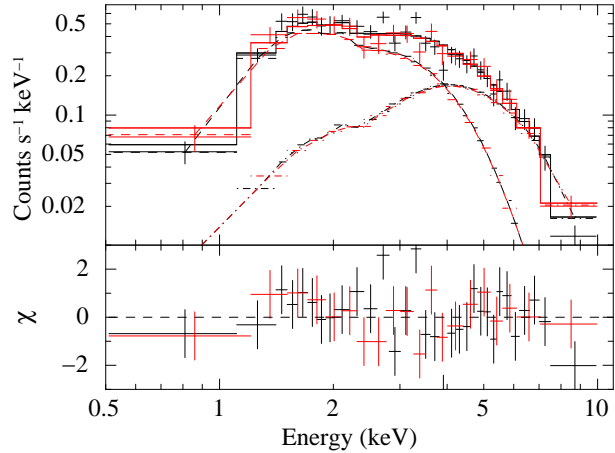}
  \caption{Swift XRT spectra outside the eclipse.
Black and red represent the 1st and 2nd observations,  respectively. 
The best-fit blackbody (dashed) $+$ CompTT (dash-dotted)
model is also shown.
Residuals between the data and the best-fit model are shown in the lower panel.}
  \label{fig:swift_spec}
\end{center}
\end{figure}

The XRT spectra of the 1st and 2nd observations outside 
the eclipse
are shown in figure~\ref{fig:swift_spec}.  
To obtain the spectrum of the 2nd observation, only the data until $30$~s before the predicted ingress were used.
The shapes of the two spectra
are quite similar to each other, and we simultaneously
analyzed the spectra to increase the photon statistics.

We tried to describe the spectra with the blackbody $+$ CompTT model.
All parameters except for the normalizations were fixed to the values 
obtained from the Chandra analysis (table~\ref{table:chandra_spec}).
The ratio between the normalizations of blackbody and of CompTT was
fixed to the value of the Chandra result.  The fit was
acceptable with $\chi^2/\nu = 1.11$~($\nu = 50$). The
observed flux was decreased from the Chandra flux by a
factor of $0.86 \pm 0.04$ for the 1st observation and 
$0.69 \pm 0.04$ for the 2nd observation.

The spectrum during the eclipse has too low statistics to be
studied in detail.  If we applied the same model as that
fitted to the spectrum outside the eclipse, the flux in the
$0.5$--$10$~keV band was $(5.1 \pm 1.9) \times
10^{-12}$~erg~cm$^{-2}$~s$^{-1}$. This value is lower than
that outside the eclipse by a factor of $(1.6 \pm 0.6)
\times 10^{-2}$.


\section{Discussion\label{section:discussion}}

\subsection{Long-term variation of the X-ray flux}
\label{subsection:flux}

\begin{figure}
\begin{center}
 \FigureFile(80mm,){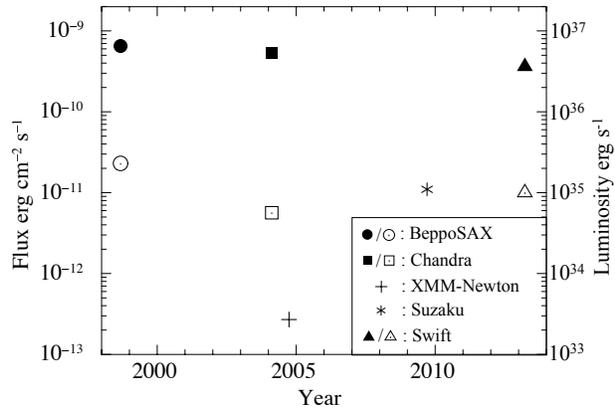}
 \caption{Observed flux of GRS~1747$-$312 in the
    $2.0$-$10$~keV band obtained in several observations,
    including the previous research
    \citep{2000A&A...355..145I}. Filled and open marks show
    the flux outside and during the eclipses, respectively.
    Vertical axis on the right hand side indicates the luminosity assuming the distance of $9.5$~kpc.}
  \label{fig:long_term_flux}
\end{center}
\end{figure}

Figure~\ref{fig:long_term_flux} shows the plot of the
observed flux in the $2.0$--$10$~keV band obtained in the
Section~\ref{subsection:Spectrum}. The BeppoSAX result
\citep{2000A&A...355..145I} is also plotted.  The flux
outside and during the eclipses are shown separately.

We can see that the flux of the outburst phase outside the eclipses is roughly
constant at $\sim 5 \times 10^{-10}$~erg~cm$^{-2}$~s$^{-1}$.
The flux during the eclipses is also constant at $\sim
10^{-11}$~erg~cm$^{-2}$~s$^{-1}$. Thus the flux during the eclipses is usually
$\sim 2$~\% of that outside the eclipses.  However, when
the XMM-Newton observed, the flux decreased further to 
$2.7 \times 10^{-13}$~erg~cm$^{-2}$~s$^{-1}$.  GRS~1747$-$312
thus shows large variability even outside
the outburst phase.

\subsection{Predicted eclipses and observations 
\label{section:discuss_lightcurve}}

Both of the Chandra and Swift light curves show ingresses at the predicted time, ${\rm MJD}
\,\,53094.16016$ ($n = 1996$) and ${\rm MJD}
\,\,56375.09967$ ($n = 8367$), respectively. In the Chandra light curve,
an egress was also detected at ${\rm MJD} \,\,53094.19020$
($n = 1996$).  Even during these eclipses, X-rays were
observed. The ratio of the flux compared to that outside the
eclipse was $\sim 2 \times 10^{-2}$. Such flux detected even during eclipses was reported in other eclipsing LMXBs; $2.7 \pm
1.0$~\% for X~1658$-$298~\citep{2001A&A...376..532O} and
$5$--$10$~\% for
EXO~0748$-$676~(\cite{2001A&A...365L.282B};
\cite{2003A&A...412..799H}).

During the Suzaku observation, two eclipses were predicted
from ${\rm MJD} \,\,55090.73879$ to ${\rm MJD}
\,\,55090.76882$ ($n = 5873$) and from ${\rm MJD
  \,\,55091.25378}$ to ${\rm MJD \,\,55051.28380}$ ($n =
5874$). However, no flux drops were seen during the
predicted times as shown with dashed lines in
figure~\ref{fig:suzaku_lc_zoom}.  
One may argue that the time stamped on the Suzaku data
was incorrect. 
We confirmed that the decrease of the 
XIS count rate of unscreened data
exactly synchronized with the SAA passages
and Earth occultations which were calculated from the satellite orbit
and attitude. We thus verified the time stamped on the XIS data,
and the vanishing of the eclipsing event is not an artifact.

Suzaku observed the sudden drop of the flux at around 60~ks
in figure~\ref{fig:suzaku_lc}, just before the type-I burst.
One may argue that the orbital phase would have been changed
for some reasons and the drop could be the eclipse.  However,
the Swift observation, which was conducted after the Suzaku
observation, confirmed that the eclipse occurred at the
predicted time. It is reasonable to consider that
the drop of the flux was different from
the eclipse event.

\begin{figure}
\begin{center}
 \FigureFile(80mm,50mm){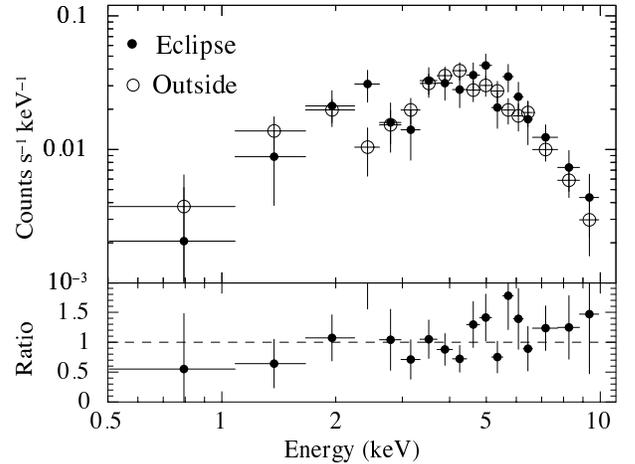}
  \caption{Suzaku spectra during the
    predicted second eclipse (filled circles) and 
    after the eclipse (open circles) obtained with XIS1.
    Ratio of the spectrum outside the eclipse 
    to that during the eclipse is shown in the lower panel.
  \label{fig:eclipse_2}
}
\end{center}
\end{figure}

Even though the flux did not change at the predicted ingresses,
some other properties might change. We then checked the
position of Source~1 during the second predicted eclipse
around $83$~ks in figure~\ref{fig:suzaku_lc_zoom}.  The
difference of the positions between outside and during the
eclipse was $(\Delta \alpha, \Delta \delta) =
(+\timeform{1."4} \pm \timeform{1."0},
+\timeform{0."2}^{+\timeform{1."3}}_{-\timeform{1."2}})$.
On the other hand, the root mean square of the fluctuation
of the Suzaku pointing direction during this observation was $\timeform{8."9}$ along
the right ascension axis, and $\timeform{5."0}$ along the
declination axis.  Thus there was no significant deference
between the source positions during and outside the
eclipse. We also checked the spectrum.
Figure~\ref{fig:eclipse_2} shows the XIS1 spectrum obtained
during the predicted second eclipse (filled circles) and that
after the predicted egress (open circles), which corresponds to
$85$--$90$~ks in figure~\ref{fig:suzaku_lc_zoom}.  We see no
significant change in the spectra.  We found that both
spectra can be described by the blackbody $+$ CompTT model
with the same parameters as those listed in the column of
``Independent fit, Interval B'' in
table~\ref{table:fit_result}. Thus neither the Suzaku image
nor spectrum showed a change coincident with the predicted
eclipse.

\subsection{Lapse of Eclipse}

We should note that the clear eclipsing events have been observed only during the outburst state. One of the possibilities to explain the lapse of the clear eclipse in the Suzaku observation is that there is a contaminant X-ray
source quite close to GRS~1747$-$312.
When GRS~1747$-$312 is not in the outburst state,
observed X-rays originate in the contaminant source
rather than GRS~1747$-$312.
When GRS~1747$-$312 is in the outburst
state, we can see the eclipse event, and the remaining X-ray
flux during the eclipse is due to the contaminant source.
This scenario can explain why the X-ray flux during the
eclipse is close to that observed with Suzaku
(figure~\ref{fig:long_term_flux}).
Based on the X-ray luminosity function and
  the radial profile of surface brightness of 47 Tuc~(
  \cite{2001Sci...292.2290G}; \cite{1995AJ....109..209G}),
  expected number of X-ray source of $>
  10^{35}$~erg~cm$^{-2}$~s$^{-1}$ within the core radius of
  a globular cluster is $\sim 0.14$. 
Since the core radius of GRS 1747$-$312 is $\timeform{3."3}$,
a contaminant source that cannot be resolved with Suzaku
is not so unlikely.

The flux of the contaminant source is mostly around 
$\sim 10^{-11}$~erg~cm$^{-2}$~s$^{-1}$
(figure~\ref{fig:long_term_flux}), which corresponds to the
luminosity of {$\sim 2 \times 10^{35}$~erg~s$^{-1}$}
assuming the distance of 9.5~kpc. The flux occasionally
decreases to a few hundredths of the usual value as
suggested by the XMM-Newton observation.
If the type-I burst observed with Suzaku 
(see figure~\ref{fig:suzaku_lc}; \cite{2014efxu.conf..158I})
came from the contaminant source,
the contaminant source should be a neutron star X-ray binary.
However, it is also possible that we saw the superposition of the contaminant source and the type-I burst occurred at GRS~1747$-$312.

If this contaminant scenario is correct, the position of the
X-ray source might change during the eclipses.  We made the
projection profiles of the Chandra image along the
declination axis during and outside the eclipse
separately. The profiles were fitted with a Gaussian
function, and we found no significant changes in the peak
position and the width of the Gaussian function.  We also
examined the Swift image during the eclipse, and the peak
position is also consistent with GRS~1747$-$312 though the
statistics are poor (see section 3.1.4). Thus we see no significant evidence for
the contaminant source in our observational data.  Note,
however, that the Chandra observation analyzed in this paper
was conducted with the CC mode and a one-dimensional image
was obtained. The other Chandra HRC-I observation did not
cover the predicted eclipse~\citep{2003A&A...406..233I}.  A
future Chandra observation covering the eclipse might be
able to detect the change of the source position.

\begin{figure}
\begin{center}
 \FigureFile(80mm,){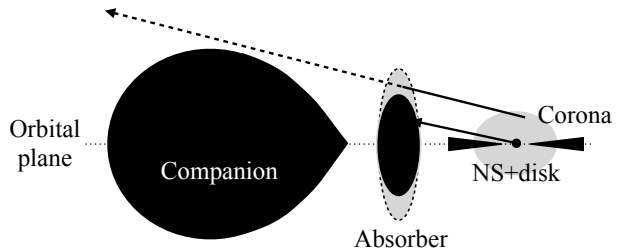}
  \caption{Schematic view of the thick material scenario
to explain the lapse of the eclipse in the Suzaku observation.
The direct X-rays from GRS~1747$-$312 is blocked by
thick material between the companion star and the neutron star.
Outer parts of the material may be less thick and patchy, 
and the Compton scattered X-rays can penetrate those parts.
\label{fig:geometry}
}
\end{center}
\end{figure}

The lapse of the eclipse during the Suzaku observation might
be explained if direct X-rays from a neutron star and an
accretion disk are perfectly blocked by thick material
between the companion star and the neutron star
(figure~\ref{fig:geometry}). 
In this condition, direct emission from near the neutron star 
that can be screened by the companion star 
cannot be seen. Then we see no change of the flux and spectrum at the predicted time of the eclipse.
 To block the direct X-rays, a column
density larger than $N_{\rm H} \geq 10^{24}$~cm$^{-2}$ is
required.  However, we saw the X-ray flux of $\sim
10^{-11}$~erg~cm$^{-2}$~s$^{-1}$ during the Suzaku
observation.  We think that the outer parts of the thick
material may have lower column density and the Compton
scattered X-rays by the hot corona can penetrate the
material. That is why the hard component of the Suzaku
spectrum suffers the heavy absorption of $\sim
10^{23}$~cm$^{-2}$ (table~\ref{table:fit_result}).  
To explain the soft component, we further assume that the outer
parts of the material is patchy. 
Then we tried to fit a partially covered model
to the Suzaku spectra.  
The CompTT model was assumed, and the ``pcfabs'' model in \texttt{XSPEC} was used to
represent the partial covering. The model can be written as
"phabs1 $\times$ pcfabs $\times$ CompTT".  The "pcfabs" is
given by
\begin{equation}
({\rm pcfabs}) = f({\rm phabs2}) + (1 - f),
\end{equation}
where $f$ represents the covering fraction. The phabs1 was
installed to represent the interstellar absorption and was
fixed to be $1.2 \times 10^{22}$~cm$^{-2}$.  The parameters
of the CompTT component were assumed to be the same for both
intervals A and B. Similar to the previous analyses, the
electron temperature was fixed to be $5.4$~keV. The fitting
was acceptable with $\chi^2/\nu = 1.07$~($\nu = 426$). 
The best-fit parameters are summarized in
table~\ref{table:pcover_fit}.
The absorption column density and covering fraction decreased from interval A to interval B,
and this can explain the soft component increased from
interval A to interval B.

The partial covering model cannot describe the XMM-Newton
spectrum with $\chi^2/\nu = 1.61 (\nu = 64)$. Therefore the
thick material scenario cannot be adopted to the low-luminosity
state generally.

\begin{table}[bhtp]
\begin{center}
\caption{Best-fit parameters of 
the partially covered CompTT model 
fitted to the Suzaku spectra.\label{table:pcover_fit}}
\begin{tabular}{lcc}
\hline
Interval & A & B \\
\hline
$N_{\rm H,pc}$\footnotemark[$*$]  (10$^{22}$~cm$^{-2} $) & $16.1^{+0.5}_{-0.5}$ & $10.5^{+0.3}_{-0.3}$ \\
Covering fraction & $0.970^{+0.007}_{-0.022}$ & $0.924^{+0.006}_{-0.006}$\\ 
$kT_{\rm seed}$~(keV) &  \multicolumn{2}{c}{$0.44^{+0.05}_{-0.05}$}\\
$kT_{\rm e}$~(keV) & \multicolumn{2}{c}{$5.4$~(fixed)}\\
$\tau$ & \multicolumn{2}{c}{$11.7^{+0.3}_{-0.2}$}\\ \hline
$\chi^2/\nu$ ($\nu$) & \multicolumn{2}{c}{$1.07$ (431)}\\
\hline
\multicolumn{3}{@{}l@{}}{\hbox to 0pt{\parbox{180mm} {\footnotesize
 \footnotemark[$*$] Hydrogen column density in units of 10$^{22}$~cm$^{-2} $.
  }\hss}}
\end{tabular}
\end{center}
\end{table}

One of the problems of the thick material scenario is that the origin of the thick material is unknown. Furthermore we see no hint of the heavy absorption during the type-I
burst at $\sim 64$~ks~(figure~\ref{fig:suzaku_lc}). The burst had a long duration with photospheric radius expansion, which indicate that the peak luminosity reached the Eddington luminosity~\citep{2014efxu.conf..158I}. Indeed, the bolometric luminosity calculated from the data at the peak was $\sim 3 \times 10^{38}$~erg~s$^{-1}$ for $9.5$~kpc, which is consistent with the Eddington luminosity for a $1.4$~M$_\odot$ neutron star. Thus the direct emission from the neutron star was seen at this moment, definitely. The thick absorber may be photo-ionized during the Type-I burst because of its high luminosity. Then the absorption column density 
can be less than that during the persistent state. However, we found no evidences for such ionization.

As discussed in section~\ref{subsubsection:suzaku_lc}, it cannot be fully rejected that weak eclipsing events were observed during the Suzaku observation.
If there was an X-ray corona of which scale height is larger than the companion star, the X-rays are not completely blocked by the companion star even during the interval of the eclipse. In this case, we could see a part of the corona emission even in the eclipse, and we would expect a weak eclipsing event. However, when we assume such a highly extended corona, the steep drop of the flux during the Suzaku observation is hard to be explained.

\section{Summary}
\label{section:summary}

GRS~1747$-$312 was observed by Chandra in 2004,
XMM-Newton in 2004, Suzaku in 2009 and Swift in 2013. 
The Chandra and Swift observations were done during the outburst. 
These observations covered the predicted time of
the eclipse, except for the XMM-Newton.  
This is the first time to study the property of GRS 1747$-$312 outside the outburst state.

During the Chandra and Swift observations, 
the eclipses were observed at the predicted time. 
Remaining X-rays were detected during the eclipses.
An averaged $2$--$10$~keV flux outside the eclipse observed
with Chandra and Swift was $\sim 5 \times 10^{-10}$~erg~cm$^{-2}$~s$^{-1}$,
but it decreased to $\sim 10^{-11}$~erg~cm$^{-2}$~s$^{-1}$ during the eclipse.
This feature was consistent with the previous research~\citep{2000A&A...355..145I} 
and some other eclipsing LMXBs.  
During the Suzaku observation, on the other hand, the eclipse was not observed.  
An averaged $2$--$10$~keV flux observed with Suzaku was $\sim 10^{-11}$~erg~cm$^{-2}$~s$^{-1}$, 
which was comparable to the flux during the eclipse.
The XMM-Newton flux was very low at $2.7 \times 10^{-13}$~erg cm$^{-2}$ s$^{-1}$. 

These results may suggest that another source is located
quite close to GRS~1747$-$312.
When GRS~1747$-$312 is not in the outburst state,
the observed X-rays came from the contaminant source
rather than from GRS~1747$-$312.
However, we did not obtain clear evidence for the contaminant source
in our data.  

The results might be explained by thick material ($N_{\rm} >
10^{24}$~cm$^{-2}$) between the neutron star and the
companion star which completely blocked the direct X-rays
from the neutron star. Outer parts of the thick material is
less thick and patchy, and X-rays Compton scattered by hot
corona can penetrate those parts. In fact, the Suzaku
spectrum can be described by the partially covered CompTT
model. The origin of the thick material, however, is not
clear.

\section*{Acknowledgments}

We thank all the Suzaku team members for their supports.  
HM is supported by
a Grants-in-Aid for Scientific Research from the MEXT, 
No. 15K13464, 24340039, and 15H02070.

We also wish to acknowledge Dr. Kazunori Ishibashi (Nagoya University)
for his helpful comments and suggestions.

\appendix
\label{section:appendix}
\section*{Suzaku HXD spectrum}


The PIN count rate was $0.65$~cts~s$^{-1}$ in the
$15$--$70$~keV band after the subtraction of the Non X-ray Background.  The
count rate of the Cosmic X-ray Background (CXB) was
negligible. On the other hand, the PIN count rate estimated
from the XIS count rate and spectrum of GRS~1747$-$312 was
$\sim\ 0.03$~cts~s$^{-1}$. The observed PIN count rate was
thus $\sim\ 20$ times larger. This suggests contamination by
other objects.

We searched other X-ray sources near GRS~1747$-$312 with the
SIMBAD
database\footnote{http://simbad.u-strasbg.fr/simbad/}.  In
the database, there is an X-ray source, IGR~J17511$-$3057 at
$(\alpha, \delta)_{\rm J2000.0}$ $=$ ($\timeform{267D.7861}$,
$\timeform{-30D.9614}$) \citep{2009ATel.2215....1N}. The
position is $\timeform{19.'4}$ north from the center of the
field of view of the PIN.  IGR~J17511$-$3057 exhibited an
outburst on 2009 September 11 (\cite{2009ATel.2196....1B};
\cite{2009ATel.2197....1M}), which was five days before the
Suzaku observation.  The count rate on September 16 of
IGR~J17511$-$3057 observed with RXTE was $\sim 70$~counts~s$^{-1}$ per a detector (Proportional Counter Unit; PCU) in the $2$--$25$~keV band~\citep{2011A&A...526A..95R}. 
Assuming a power-law spectrum with $\Gamma = 2.0$, 
the count rate corresponds to $\sim 0.52$~cts~s$^{-1}$ of PIN.  
We thus concluded that the PIN count rate was dominated by 
IGR~J17511$-$3057, 
and we decided not to use the PIN data for our analysis.

\bigskip

\end{document}